\documentclass[10pt]{iopart}
\usepackage{graphicx}
\usepackage{wrapfig}
\usepackage{iopams}
\usepackage{setspace}
\usepackage{cite}
\usepackage{hyperref}

\usepackage{color}

\newcommand{\text}[1]{#1}

\begin{document}

\title[Simulations of fast ion wall loads in AUG...]{Simulations of fast ion wall loads in ASDEX Upgrade in the presence of magnetic perturbations due to ELM mitigation coils}
\author{O. Asunta$^1$, S. {\"A}k{\"a}slompolo$^1$, T. Kurki-Suonio$^1$, T. Koskela$^1$, S. Sipil{\"a}$^1$, A. Snicker$^1$, M. Garc{\'i}a-Mu{\~n}oz$^2$, and the ASDEX Upgrade team}
\address{$^1$ Aalto University, Euratom-Tekes Association, P.O. Box 14100, FI-00076 AALTO, Finland}
\address{$^2$ Max-Planck-Institut f{\"u}r Plasmaphysik, EURATOM Association, Boltzmannstr. 2, D-85748, Garching, Germany}
\ead{otto.asunta@aalto.fi}

\begin{abstract}

The effect of ASDEX Upgrade (AUG) ELM mitigation coils on fast ion wall loads was studied with the fast particle following Monte Carlo code ASCOT. Neutral beam injected (NBI) particles were simulated in two AUG discharges both in the presence and in the absence of the magnetic field perturbation induced by the eight newly installed in-vessel coils. In one of the discharges (\#26476) beams were applied individually, making it a useful basis for investigating the effect of the coils on different beams. However, no ELM mitigation was observed in \#26476, probably due to the low plasma density. Therefore, another discharge (\#26895) demonstrating clear ELM mitigation was also studied.
The magnetic perturbation due to the in-vessel coils has a significant effect on the fast particle confinement, but only when total magnetic field, $B_{\textrm{tot}}$, is low.
When $B_{\textrm{tot}}$ was high, the perturbation did not increase the losses, but merely resulted in redistribution of the wall power loads. Hence, it seems to be possible to achieve ELM mitigation using in-vessel coils, while still avoiding increased fast ion losses, by simply using a strong  $B_{\textrm{tot}}$.
Preliminary comparisons between simulated and experimental Fast Ion Lost Detector (FILD) signals show a reasonable correspondence. 
\footnote{\underline{O. Asunta et al 2012 Nucl. Fusion {\bf 52} 094014}, This is an author-created, un-copyedited version of an article accepted for publication in Nuclear Fusion. IOP Publishing Ltd is not responsible for any errors or omissions in this version of the manuscript or any version derived from it. The Version of Record is available online at \url{http://dx.doi.org/10.1088/0029-5515/52/9/094014}.}

\end{abstract}

\maketitle

%
\section{Introduction}
\begin{figure}[b]
\begin{center}
\includegraphics[width=0.70\textwidth]{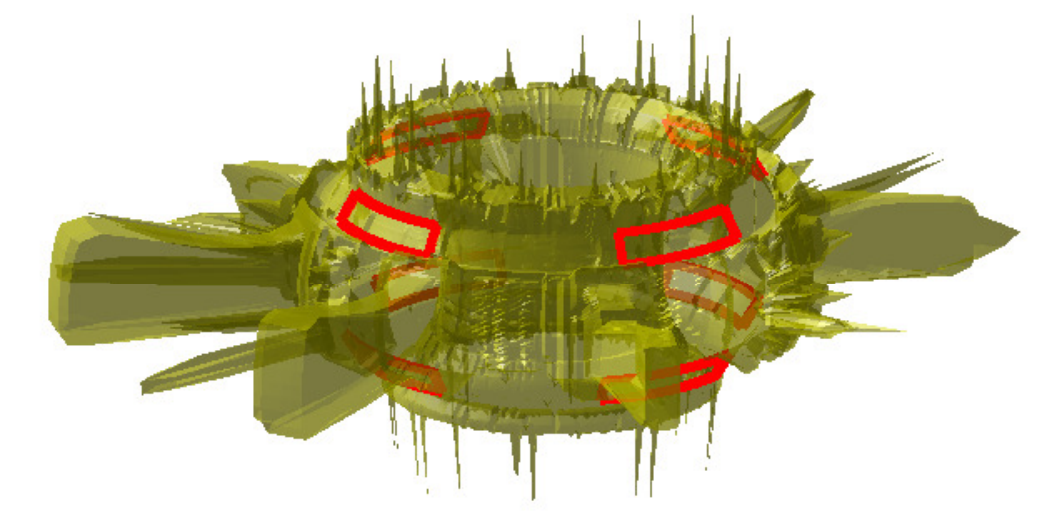}
\end{center}
\caption{The 3D-wall structure of ASDEX Upgrade used in ASCOT. The in-vessel coils in operation are depicted in red.}
\label{fig:aug3d_w_coils}
\end{figure}

\noindent Mitigation of edge localized modes (ELMs) is vital for successful high-confinement mode (H-mode) operation of ITER~\cite{Federici06_PWI_in_ITER}. 
At DIII-D, magnetic field perturbations were found to reduce the size and increase the frequency of ELMs without deteriorating the core plasma performance~\cite{Evans04_ELM_suppression_DIII-D}. Experiments aiming at ELM mitigation using magnetic perturbations have since been performed on various tokamaks, e.g. DIII-D~\cite{Evans08_RMP_ELM_suppression_DIII-D,Petty10_ELM_suppression_DIII-D}, JET~\cite{Liang07_ELM_coils_JET} and MAST~\cite{Kirk11_ELM_coils_MAST}.
Recently also ASDEX Upgrade (AUG) was furnished with in-vessel saddle coils in order to study ELM mitigation~\cite{Suttrop09_in-vessel_coils_AUG}. By the time of this work, eight out of the designed twenty-four coils had already been installed and their locations are shown in Fig.~\ref{fig:aug3d_w_coils}. %
Running a current in the coils in the positive (negative) direction creates a magnetic field mainly in outward (inward) radial direction. First experiments using the coils showed clear mitigation of ELMs without compromising the plasma performance (e.g. stored energy and pedestal top density)~\cite{suttrop11_ELM_mitigation_AUG}.

While the magnetic perturbation created by the in-vessel coils has been found to have the desirable effect on ELMs, it might be harmful for the fast ion confinement. Indeed, local perturbations, e.g. the one due to tritium breeding test blanket modules (TBMs) projected for ITER, have been found to cause increased and more localized fast ion losses~\cite{ascot_wall2009}.

This work investigates how the magnetic perturbation created by the in-vessel coils affects the confinement and losses of fast particles. Neutral beam injected (NBI) particles were simulated in AUG discharges \#26476 and \#26895 in the presence and absence of the said magnetic perturbation. In discharge \#26476, ELM mitigation was not observed. It was chosen because beams were turned on/off one at a time, which makes the discharge ideal for studying the effect of the in-vessel coils on individual beams. Discharge \#26895, on the other hand, is a typical example of successful ELM mitigation~\cite{Suttrop11_ELM_mitigation_studies}. The simulations were done with the test particle orbit following Monte Carlo code ASCOT~\cite{ascot_wall2009,Heikkinen1993215}. Results from a synthetic diagnostic in ASCOT were compared with those of the fast ion loss detector (FILD)~\cite{garcia-munoz:053503}.

The structure of the paper is as follows: ASCOT-code and the simulated discharges are presented in Sec.~\ref{sec:ascotSimulations}. Section~\ref{sec:wallLoads} discusses the changes in wall loads induced by the in-vessel coils, whereas in Sec.~\ref{sec:fild} the simulations are compared with FILD measurements. Finally, in Sec.~\ref{sec:conclusions},  conclusions are drawn and future work is discussed.%

\section{ASCOT simulations}
\label{sec:ascotSimulations}
\subsection{Simulated cases}
\noindent ASCOT~\cite{ascot_wall2009,Heikkinen1993215} is able to take into account the full 3D structures of both the magnetic field and the first wall of the device. This makes it an ideal tool for modelling fast ion wall loads, particularly in non-axisymmetric magnetic fields. In this work, the most recent 3D wall structure of AUG (see Fig.~\ref{fig:aug3d_w_coils}), updated to include the modifications for the 2010--2011 experimental campaign, and the magnetic fields from AUG discharges \#26476 ($B_{\textrm{t}}=1.8$~T, $I_{\textrm{p}}=0.8$~MA), and \#26895 ($B_{\textrm{t}}=2.5$~T, $I_{\textrm{p}}=0.8$~MA) were used. The toroidal field ripple as well as the perturbation due to the in-vessel coils were calculated using the vacuum field approximation. That is, due to the lack of any reliable calculations plasma shielding, that in reality reduces the effect of the perturbation, was not taken into account. Consequently, the fast ion losses given in this work represent the 'worst case scenario'.

\begin{figure}
\begin{center}
\includegraphics[width=0.40\textwidth]{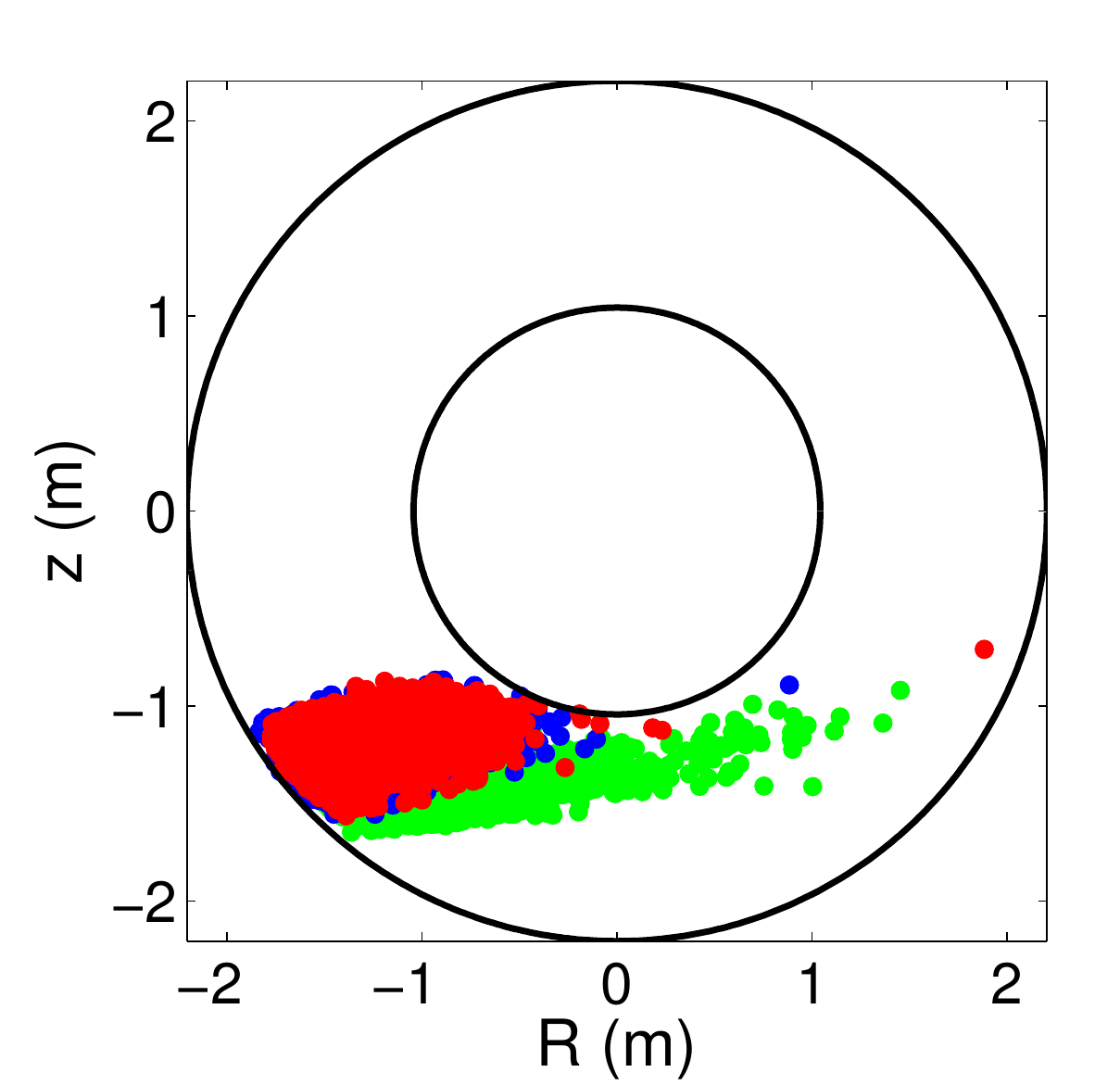}
\hspace{0.8cm}
\includegraphics[width=0.25\textwidth]{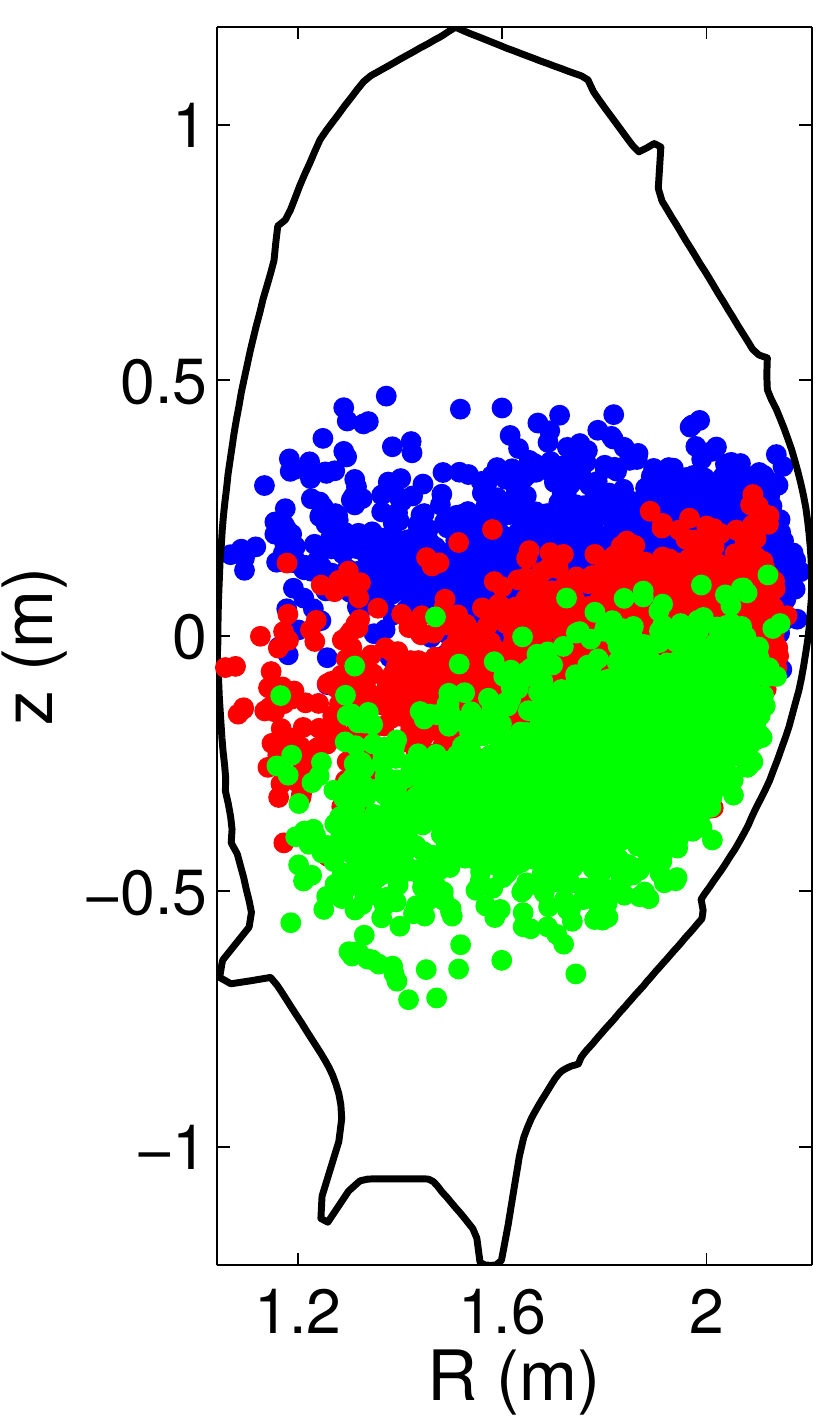}
\end{center}
\caption{The beams used in AUG discharge \#26476 illustrated with corresponding test particle ensembles. Beams Q5 (blue, horizontal perpendicular), Q6 (green, downward parallel), and Q8 (red, downward perpendicular) are viewed from above (left) and in poloidal cross-section (right).}
\label{fig:beams}
\end{figure}

In \#26476, six different cases were studied; three neutral beams Q5, Q6, and Q8 (93~keV and 2.5~MW each) were run individually, each with both $I_{\textrm{coil}}=0.0$~A, and $I_{\textrm{coil}}=\pm 0.95$~kA current in the in-vessel coils. The beams used in \#26476 are illustrated with corresponding test particle ensembles in Fig.~\ref{fig:beams}. The purpose of simulating also discharge \#26895 was to study a typical ELM mitigation discharge. There, one 60~keV (Q3) and two 93~keV (Q6 and Q8) beams, each providing 2.5~MW of heating power, were applied simultaneously. During the discharge, the current in the in-vessel coils was switched from $I_{\textrm{coil}}=0.0$~A to $I_{\textrm{coil}}=\pm 0.96$~kA. 

\begin{figure}
\begin{center}
\includegraphics[width=0.35\textwidth]{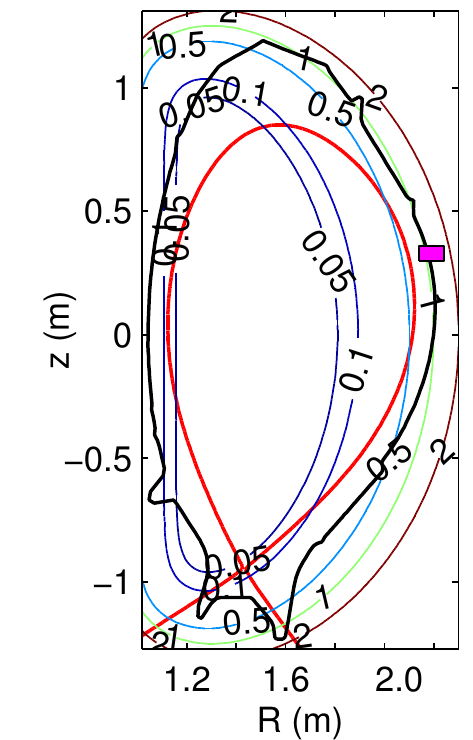}
\hspace{-1.cm} \raisebox{6.5cm}{\text{(a)}} \hspace{-0.1cm}
\includegraphics[width=0.35\textwidth]{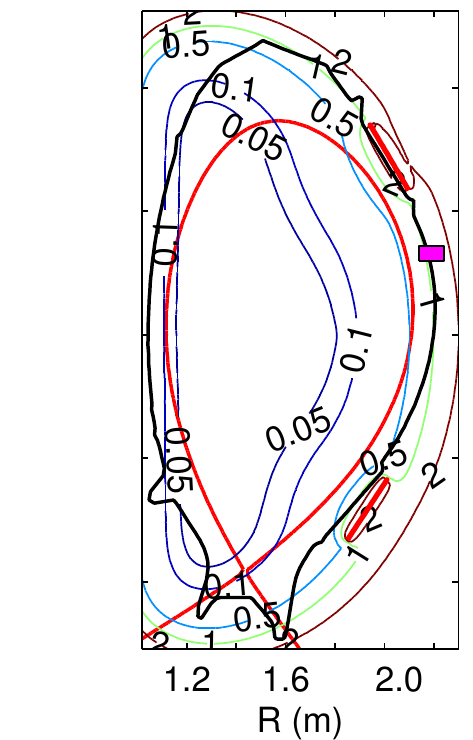}
\hspace{-1.cm} \raisebox{6.5cm}{\text{(b)}} %
\end{center}
\caption{Ripple maps depicting $\delta = 100 \times \frac{B_{\textrm{max}}-B_{\textrm{min}}}{B_{\textrm{max}}+B_{\textrm{min}}}$ in \#26476 with (a) $I_{\textrm{coil}}=0.0$~A, i.e. toroidal field ripple only, and (b) $I_{\textrm{coil}}=0.95$~kA current in the in-vessel coils. The coils are indicated by red bars on the outboard side of (b) and the FILD location is marked by the magenta square.}
\label{fig:rippleMap26476}
\end{figure}

\begin{figure}
\begin{center}
\includegraphics[width=0.35\textwidth]{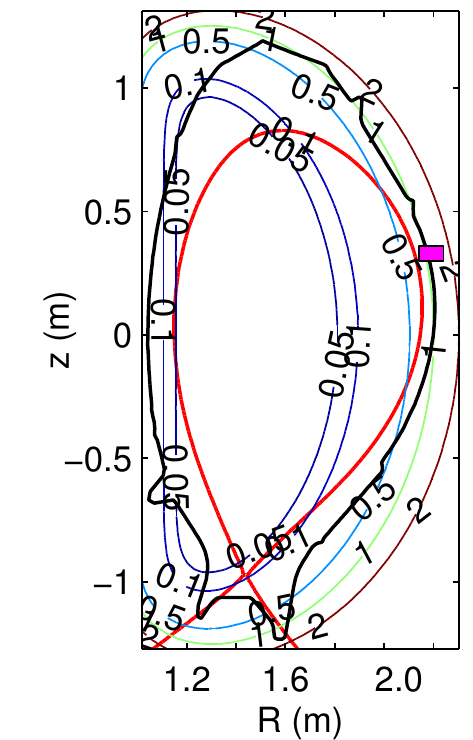}
\hspace{-1.cm} \raisebox{6.5cm}{\text{(a)}} \hspace{-0.1cm}
\includegraphics[width=0.35\textwidth]{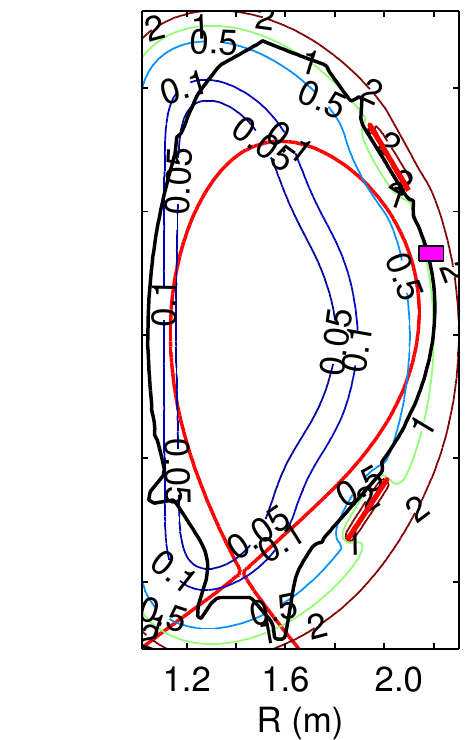}
\hspace{-1.cm} \raisebox{6.5cm}{\text{(b)}}
\end{center}
\caption{Ripple maps depicting $\delta = 100 \times \frac{B_{\textrm{max}}-B_{\textrm{min}}}{B_{\textrm{max}}+B_{\textrm{min}}}$ in \#26895 with (a) $I_{\textrm{coil}}=0.0$~A, i.e. toroidal field ripple only, and (b) $I_{\textrm{coil}}=0.96$~kA current in the in-vessel coils. The coils are indicated by red bars on the outboard side of (b) and the FILD location is marked by the magenta square.}
\label{fig:rippleMap26895}
\end{figure}

In both simulated discharges, the coils were used in the odd parity configuration (i.e. opposite polarity of upper and lower coils), creating an $n=2$ perturbation. Figures.~\ref{fig:rippleMap26476} and~\ref{fig:rippleMap26895} show the ripple maps for the toroidal field ripple only (a), and toroidal field ripple together with the effect of the in-vessel coils (b). It is worth mentioning that with the larger $B_{\textrm{t}}$ (2.5~T in \#26895 compared to 1.8~T in \#26476) the coils have a remarkably smaller effect on the total magnetic field. This is apparent from bump in the light blue $\delta=0.5~\%$ contour in front of the upper set of coils, that is more noticeable in Fig.~\ref{fig:rippleMap26476}(b) than in Fig.~\ref{fig:rippleMap26895}(b). The resulting differences in flux surface deformation turn out to be very crucial for fast particle losses.


\begin{figure}
\begin{center}
\includegraphics[width=0.45\textwidth]{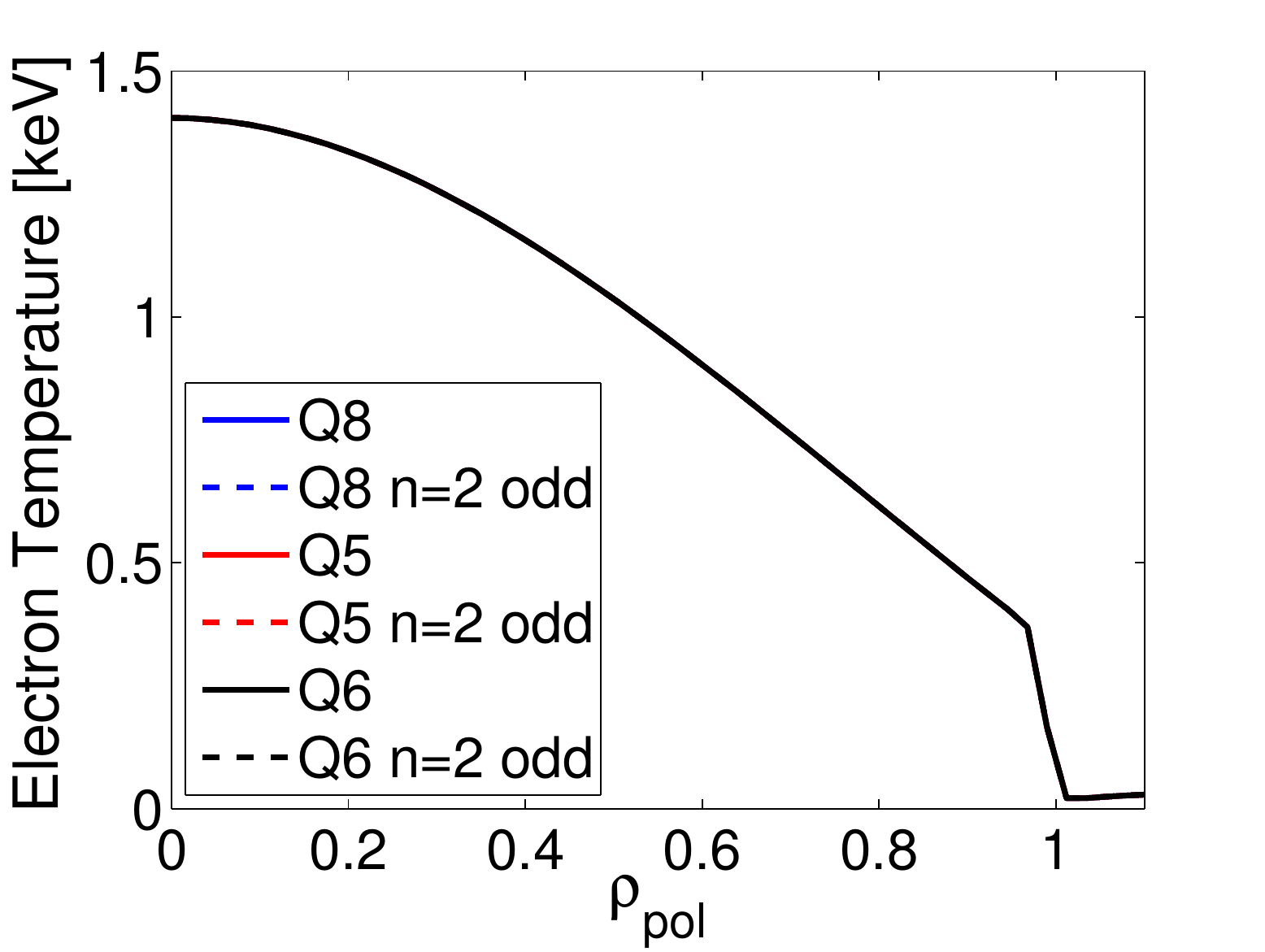}
\hspace{-1.6cm} \raisebox{3.6cm}{\text{(a)}} \hspace{1.0cm}
\includegraphics[width=0.45\textwidth]{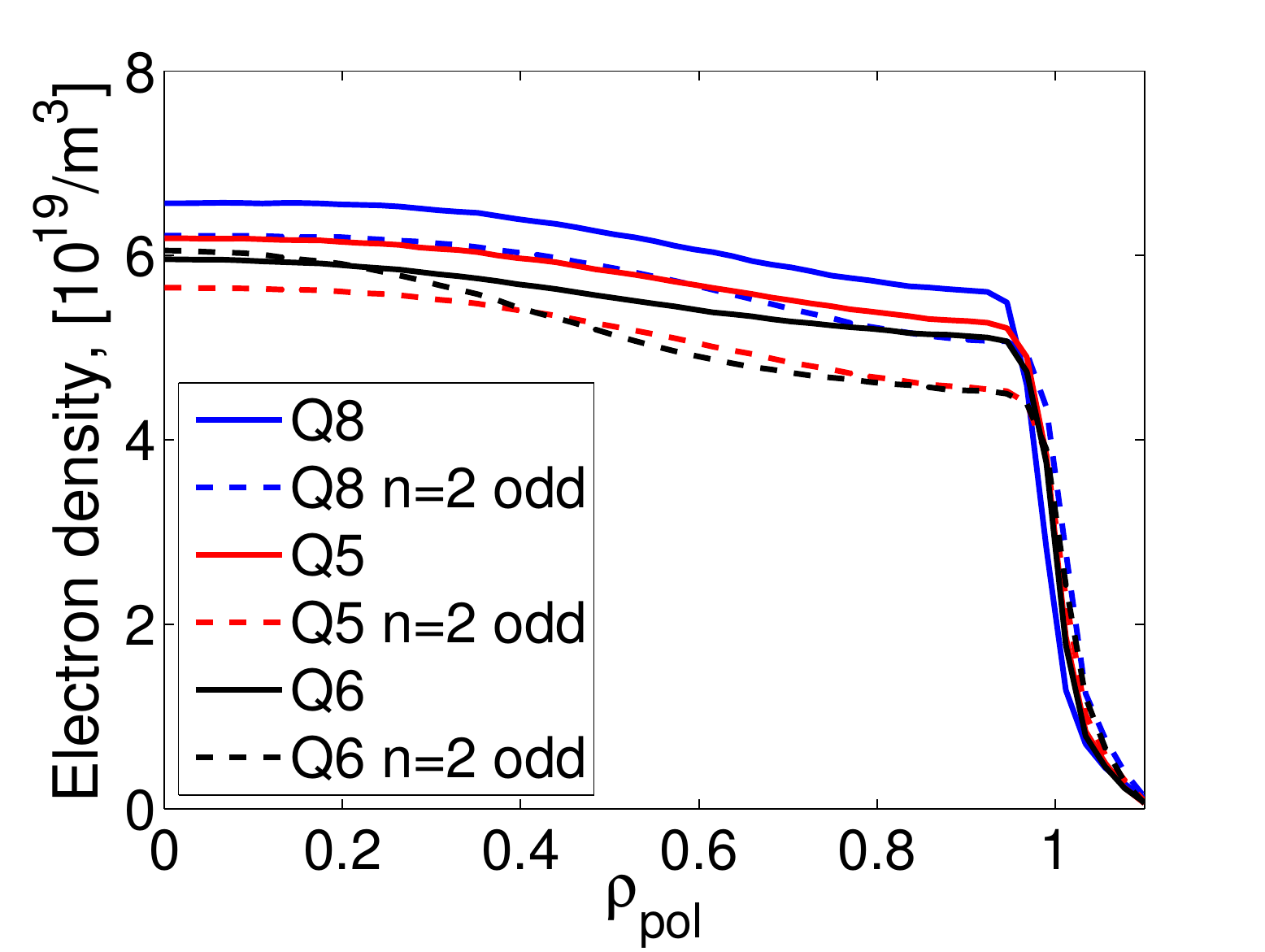}
\hspace{-1.6cm} \raisebox{3.6cm}{\text{(b)}}
\end{center}
\caption{(a) Temperature and (b) density profiles used for the six simulated cases for discharge \#26476.}
\label{fig:26476_tn}
\end{figure}

\begin{figure}
\begin{center}
\includegraphics[width=0.45\textwidth]{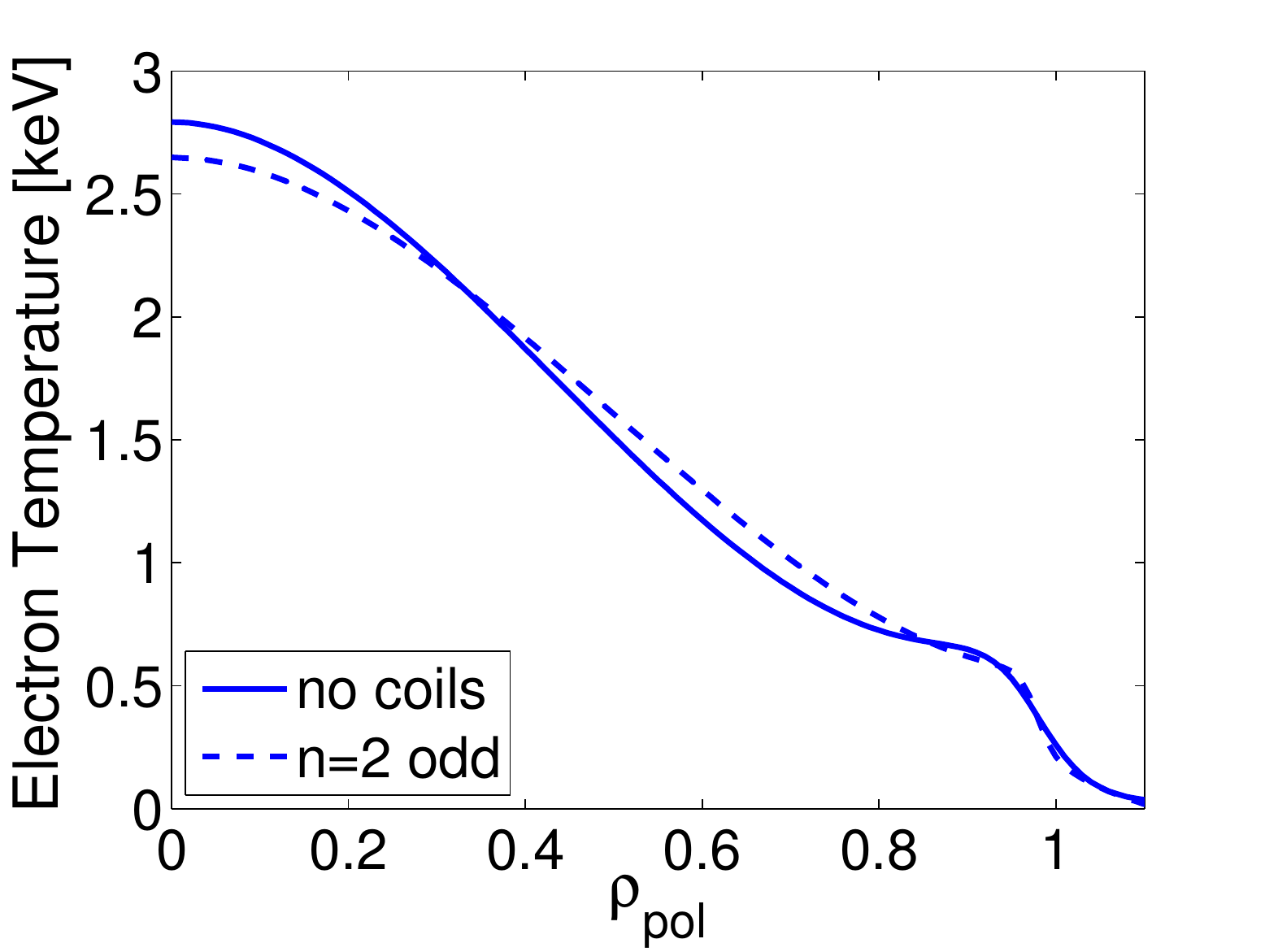}
\hspace{-1.6cm} \raisebox{3.6cm}{\text{(a)}} \hspace{1.0cm}
\includegraphics[width=0.45\textwidth]{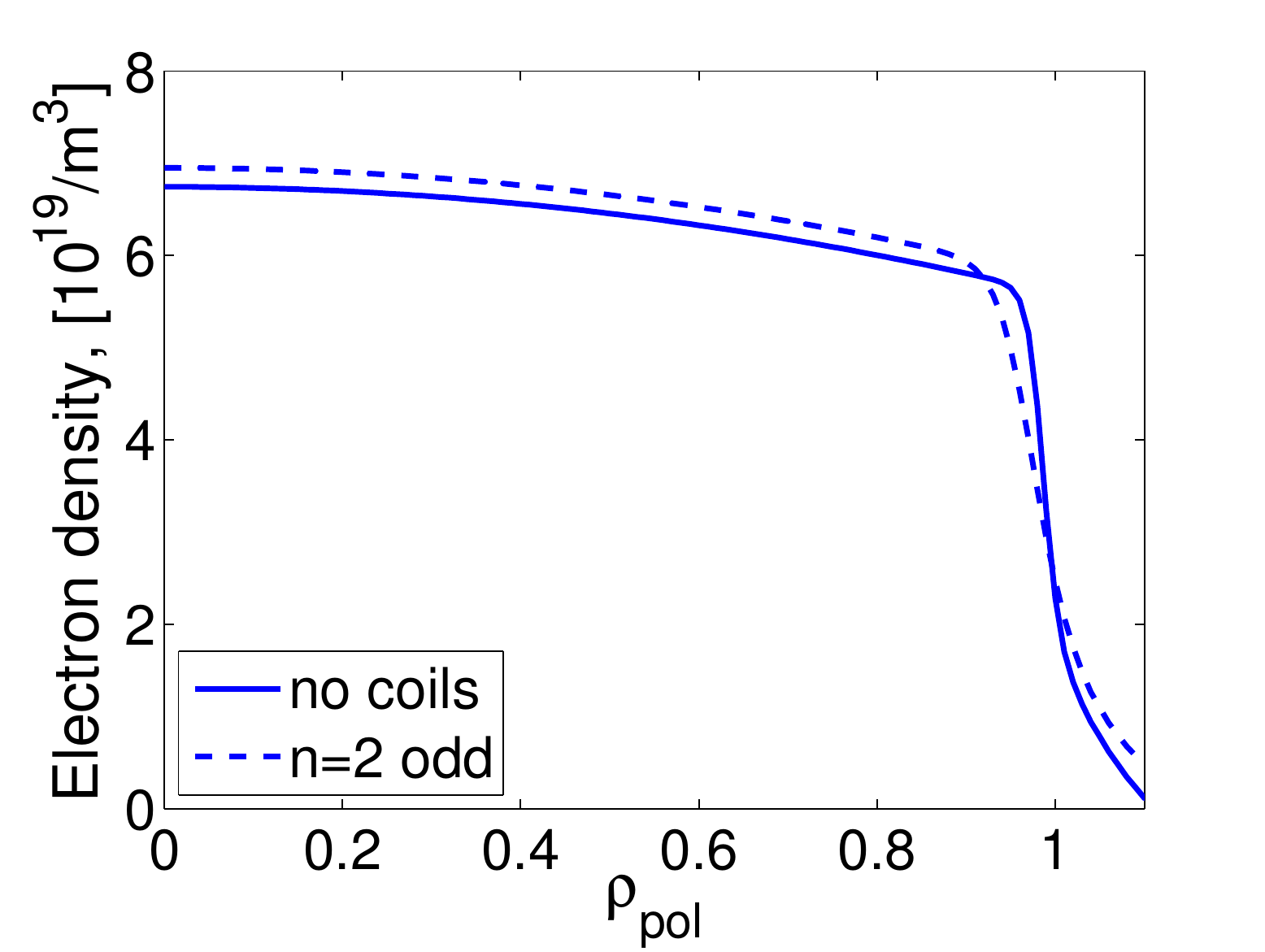}
\hspace{-1.6cm} \raisebox{3.6cm}{\text{(b)}}
\end{center}
\caption{(a) Temperature and (b) density profiles used in simulations for discharge \#26895 at $t=2.25$~s ($I_{\textrm{coil}}=0.0$~A) and $t=3.11$~s ($I_{\textrm{coil}}=0.96$~kA, $n=2$ odd). The change in temperature and density induced by switching on the coils is minimal.}
\label{fig:26895_tn}
\end{figure}

The density and temperature profiles for the six cases in \#26476 were obtained with IDA~\cite{IDA_Fischer10} (Integrated Data Analysis). IDA is a tool that uses Bayesian probability theory to coherently combine profile data from different diagnostics. It offers a consistent way of handling diagnostic data and also provides systematic and unified error evaluation. The resulting profiles for \#26476 are presented in Fig.~\ref{fig:26476_tn}. The electron temperature measurements for all the cases were within each others' uncertainty of measurement and, therefore, a constant $T_{e}$ profile was used. In the simulations it was further assumed that $T_{i}=T_{e}$. The variation in electron density between the six cases was also very small but, since it seemed to have a clear trend (i.e. the density dropped when the coil current was on), this variation was taken into account in the simulations by using different profiles for each case. Sensitivity analysis did, however, show that the results are not significantly affected by the small variations in electron density. Instead, the differences in wall loads between cases with coils on/off are dominated by the changes in magnetic configuration caused by the in-vessel coils. The temperature and density profiles for the typical ELM mitigation discharge \#26895 are presented in Fig.~\ref{fig:26895_tn}. Again, switching on the coils caused only minor changes in the profiles.

A quasineutral plasma of deuterium with very little boron, resulting in $Z_{\textrm{eff}}=1.05$, was assumed in all the simulations. Z-effective was not expected to play a significant role in the simulations presented here. To verify this, also a quasineutral plasma with nitrogen impurity and $Z_{\textrm{eff}}=1.3$ was tested. As predicted, this small a change in $Z_{\textrm{eff}}$ had no effect on the results.

\subsection{Orbit following}

For every simulation, 800000 NBI test particles were generated using ASCOT NBI~\cite{AsuntaEFTC2009}. 
The particles' orbits were then traced until they either hit a material surface or had cooled down to two and half times the local energy of thermal ions. To save computational time, while inside the separatrix, the particle guiding-centers (GC) were followed. However, when the guiding-center time step line segment came within the distance of one Larmor radius from the first wall, the method of particle following was switched from guiding-center to full-orbit (FO) following, and the step retaken~\cite{5545410}. The purpose of this switch was to accurately resolve the power load pattern on the walls of the device. In the switch, the particle was assigned a random gyro-phase, while keeping all physical variables constant, and then followed until it hit the wall. The effect and importance of combining the full-orbit wall collision model to the guiding-center following (GC+FO) was tested by comparing the results with a simulation where particle was deemed to hit the wall only when its GC crossed the wall surface.

ASCOT is also capable of following the particles' full orbit (FO) throughout the simulation~\cite{SipilaEPS2010}. This is, however, computationally much (i.e. tens of times) more expensive than the GC+FO method described above, and should therefore be used only when absolutely necessary. To check the validity of the GC+FO results, the parallel beam Q6 in \#26476 was also simulated using the full-orbit following and a limited number (20000) test particles.

\subsection{Fast Ion Loss Detector}
The fast ion loss detector (FILD) at AUG consists of a scintillator plate protected by a cylindrical graphite casing~\cite{garcia-munoz:053503}. A carefully designed collimator slit in the casing lets particles with certain pitch angle and Larmor radius to enter the probe and hit the scintillator plate. Particles with different energy and pitch will hit different parts of the plate. An optical telescope is then used to record the light emissions from the plate.

To achieve maximal realism, ASCOT could model the casing, the collimator, and the plate. However, due to the small size of the collimator slit, only a tiny fraction of the test particles would actually hit the plate. Therefore, the pitch and energy distribution of all particles that hit the casing are statistically analysed. The model for the casing is a cylinder with radius of 0.04~m located at $R=2.14...2.24$~m, $z=0.33$~m.
\section{Simulated wall loads due to in-vessel coils}
\label{sec:wallLoads}
\begin{figure}[tb]
\begin{minipage}{0.99\linewidth}
\includegraphics[width=0.341\linewidth]{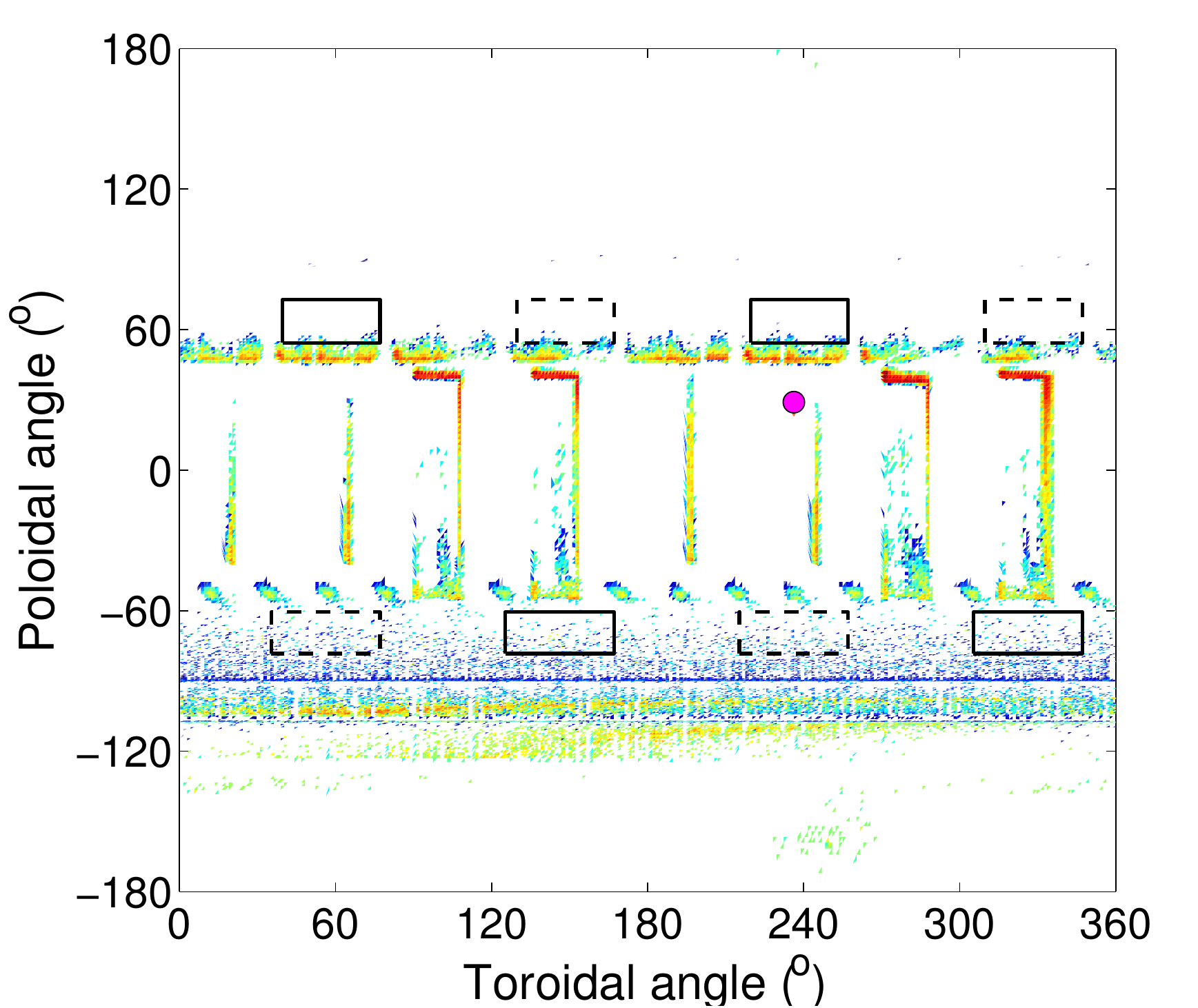}
\hspace{-3.9cm} \raisebox{3.1cm}{\text{(a)}} \hspace{2.9cm} 
\includegraphics[width=0.34\linewidth]{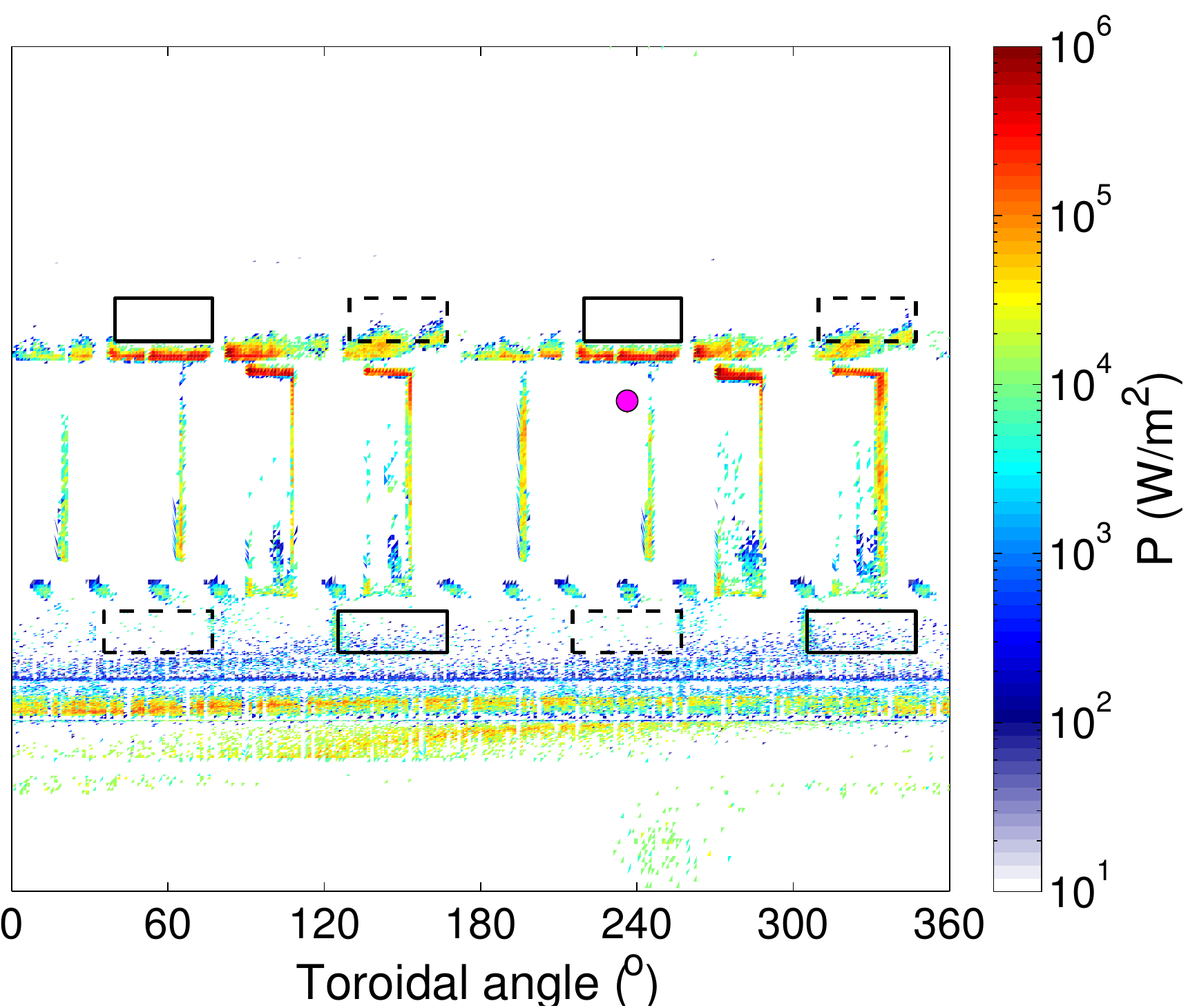}
\hspace{-4.4cm} \raisebox{3.1cm}{\text{(b)}} \hspace{3.4cm} 
\includegraphics[width=0.328\linewidth]{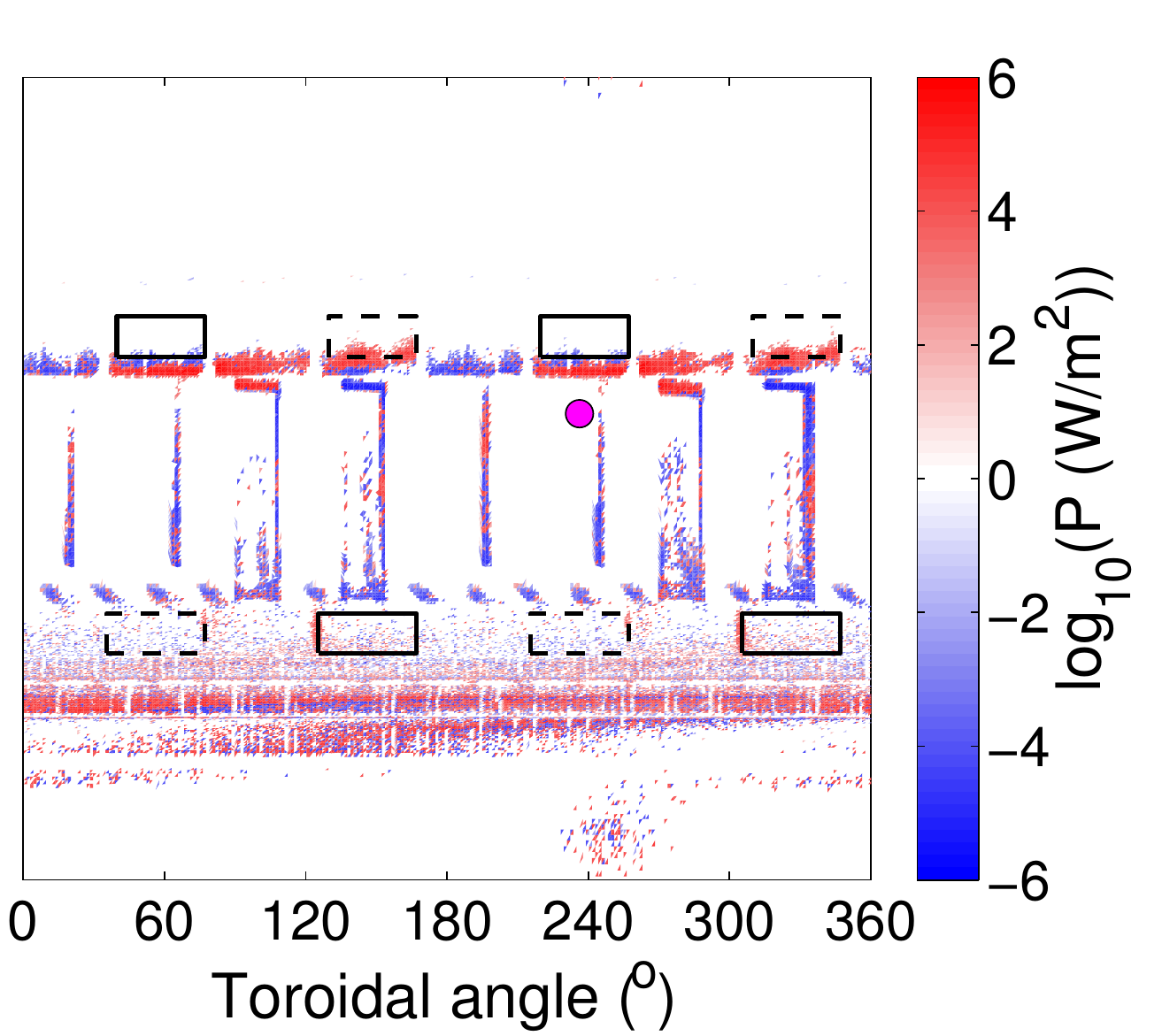}
\hspace{-4.2cm} \raisebox{3.1cm}{\text{(c)}}\\ 
\includegraphics[width=0.341\linewidth]{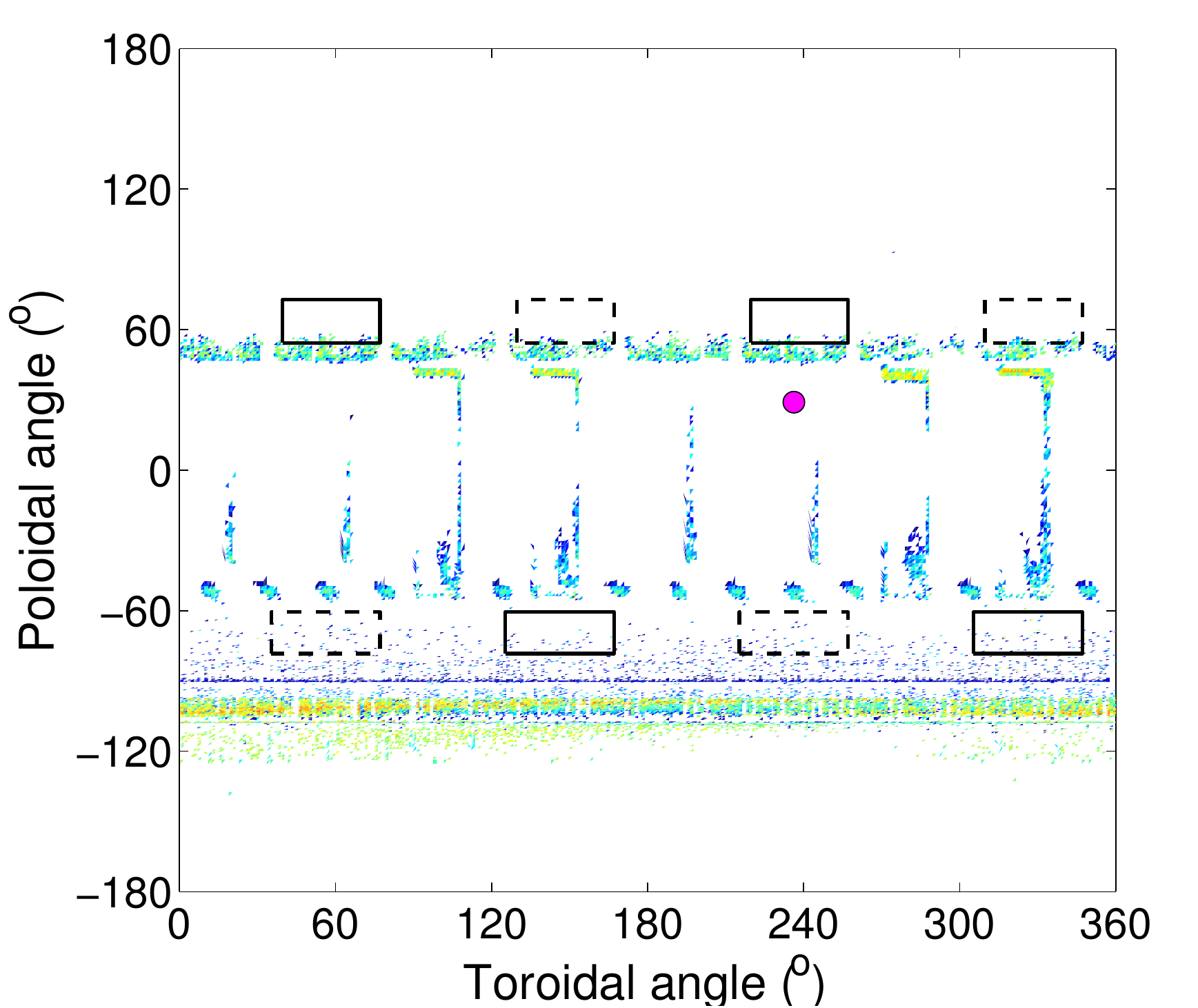}
\hspace{-3.9cm} \raisebox{3.1cm}{\text{(d)}} \hspace{2.9cm} 
\includegraphics[width=0.34\linewidth]{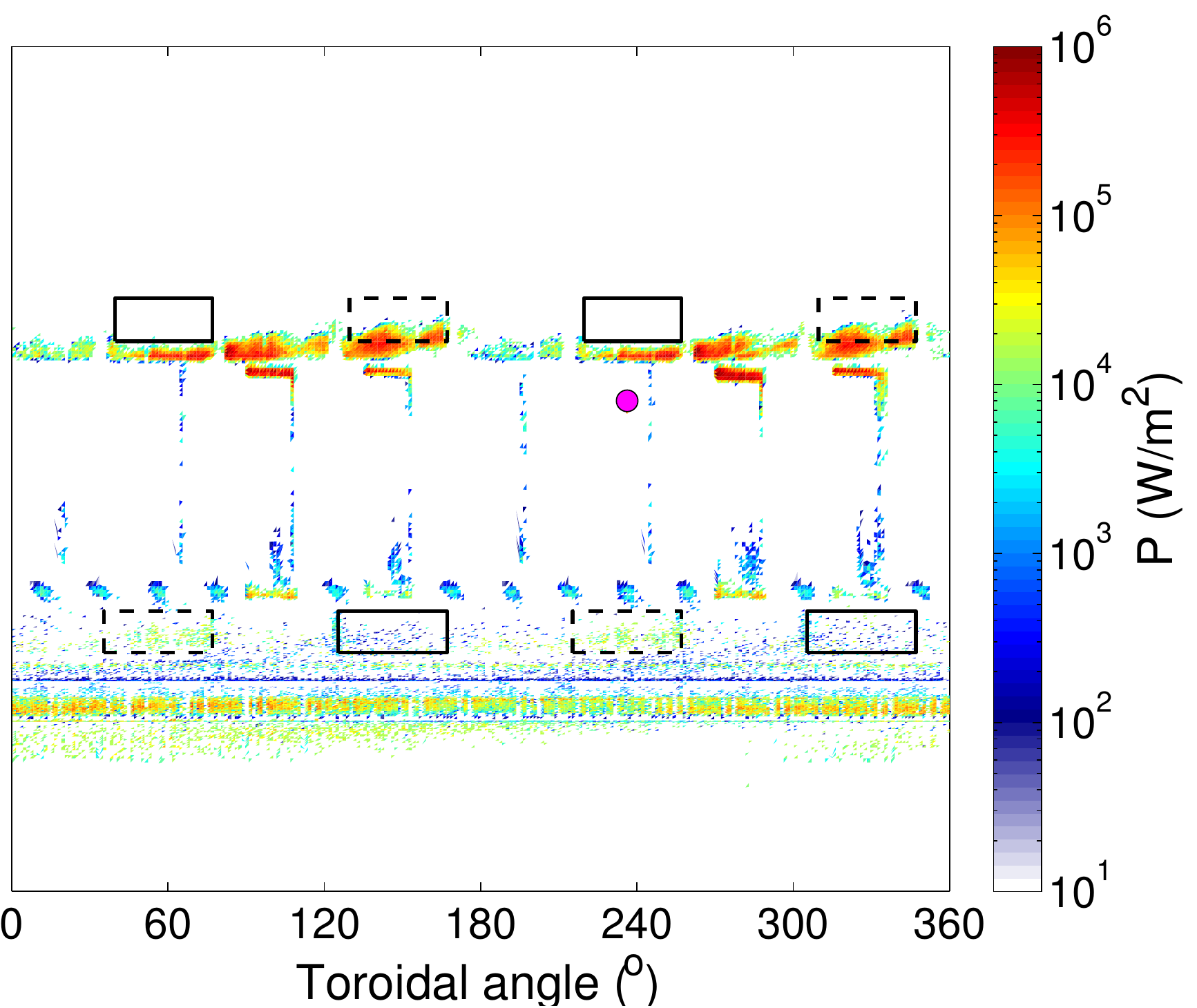}
\hspace{-4.4cm} \raisebox{3.1cm}{\text{(e)}} \hspace{3.4cm}
\includegraphics[width=0.328\linewidth]{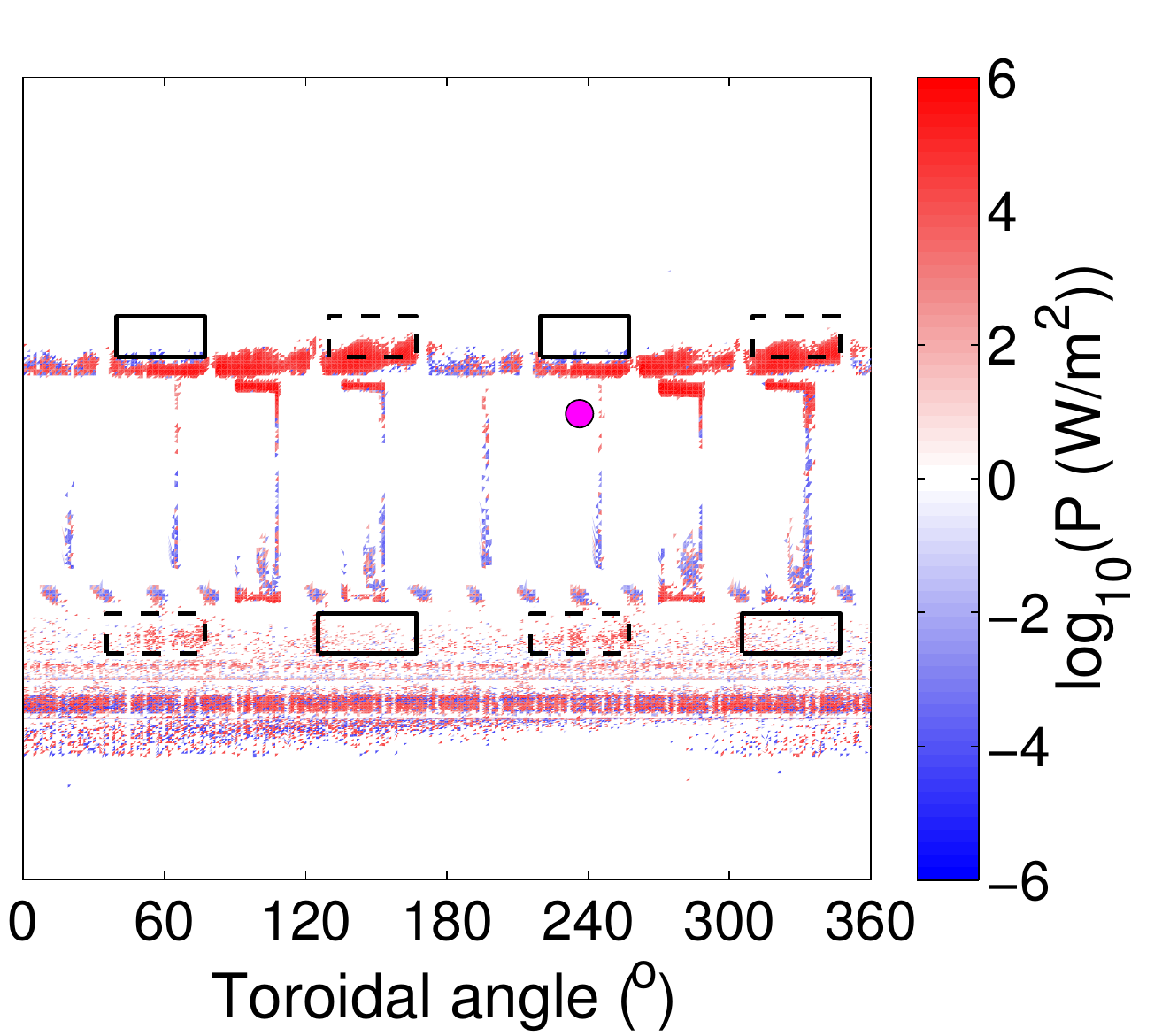}
\hspace{-4.2cm} \raisebox{3.1cm}{\text{(f)}}\\ 
\includegraphics[width=0.341\linewidth]{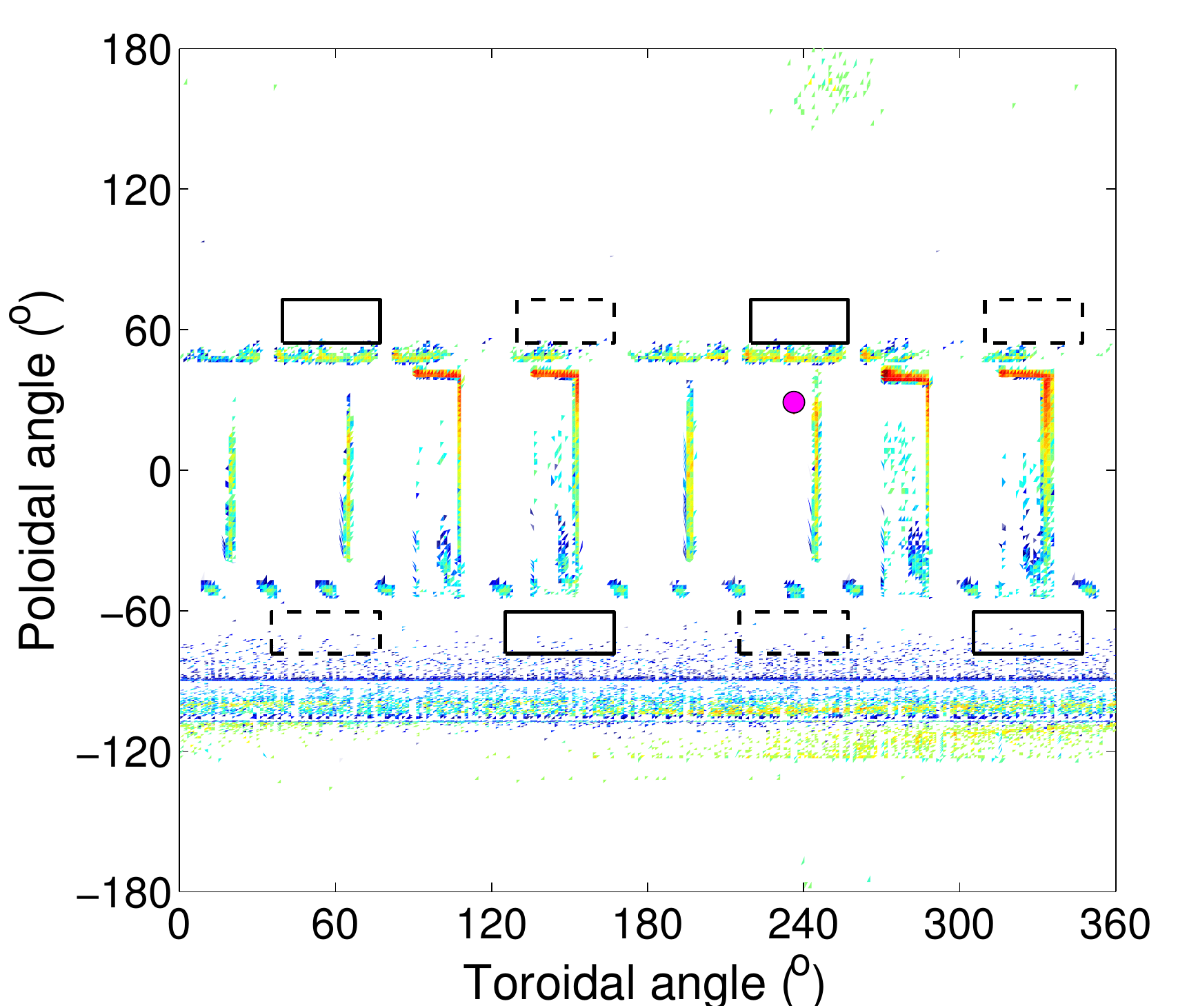}
\hspace{-3.9cm} \raisebox{3.1cm}{\text{(g)}} \hspace{2.9cm} 
\includegraphics[width=0.34\linewidth]{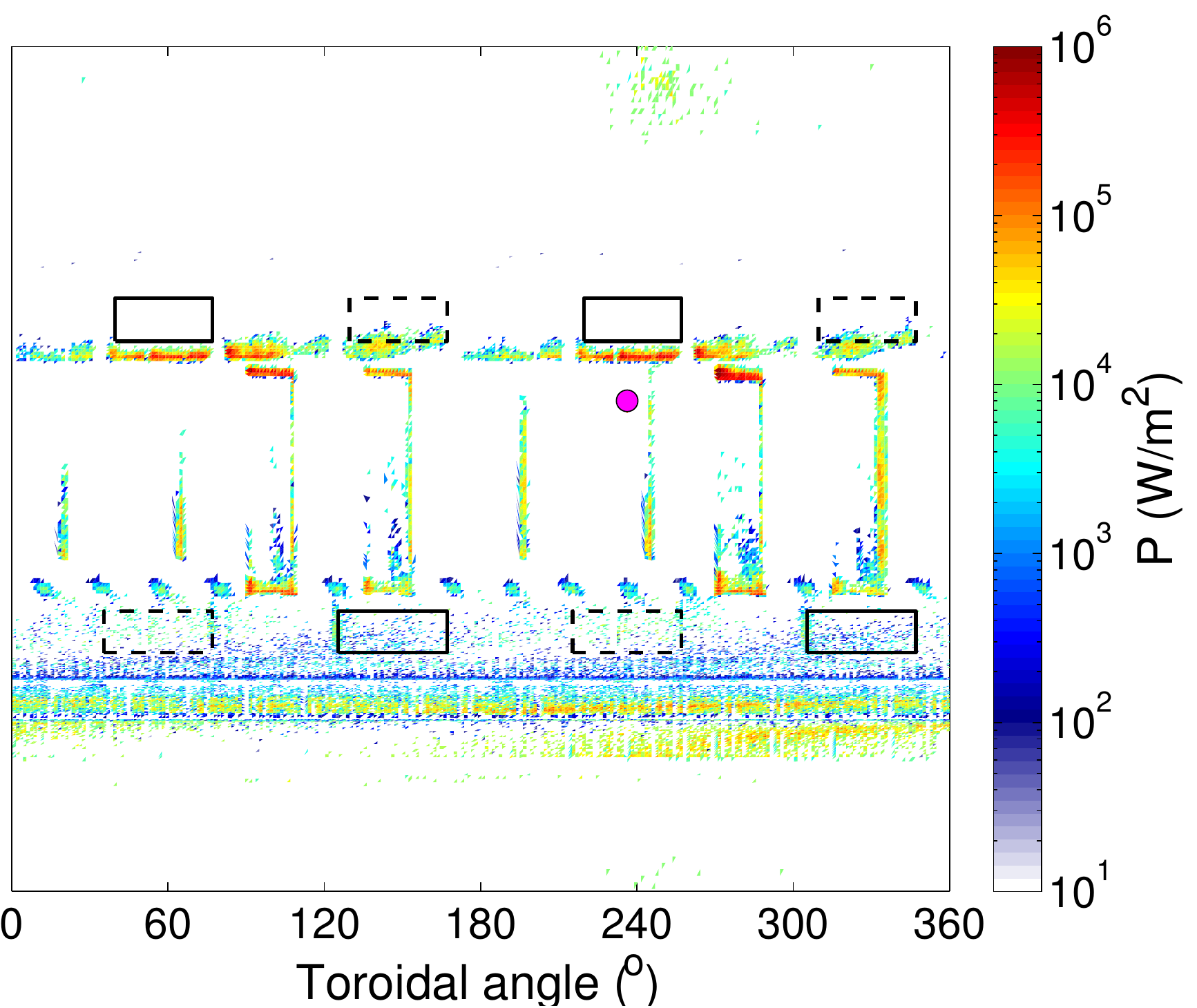}
\hspace{-4.4cm} \raisebox{3.1cm}{\text{(h)}} \hspace{3.4cm}
\includegraphics[width=0.328\linewidth]{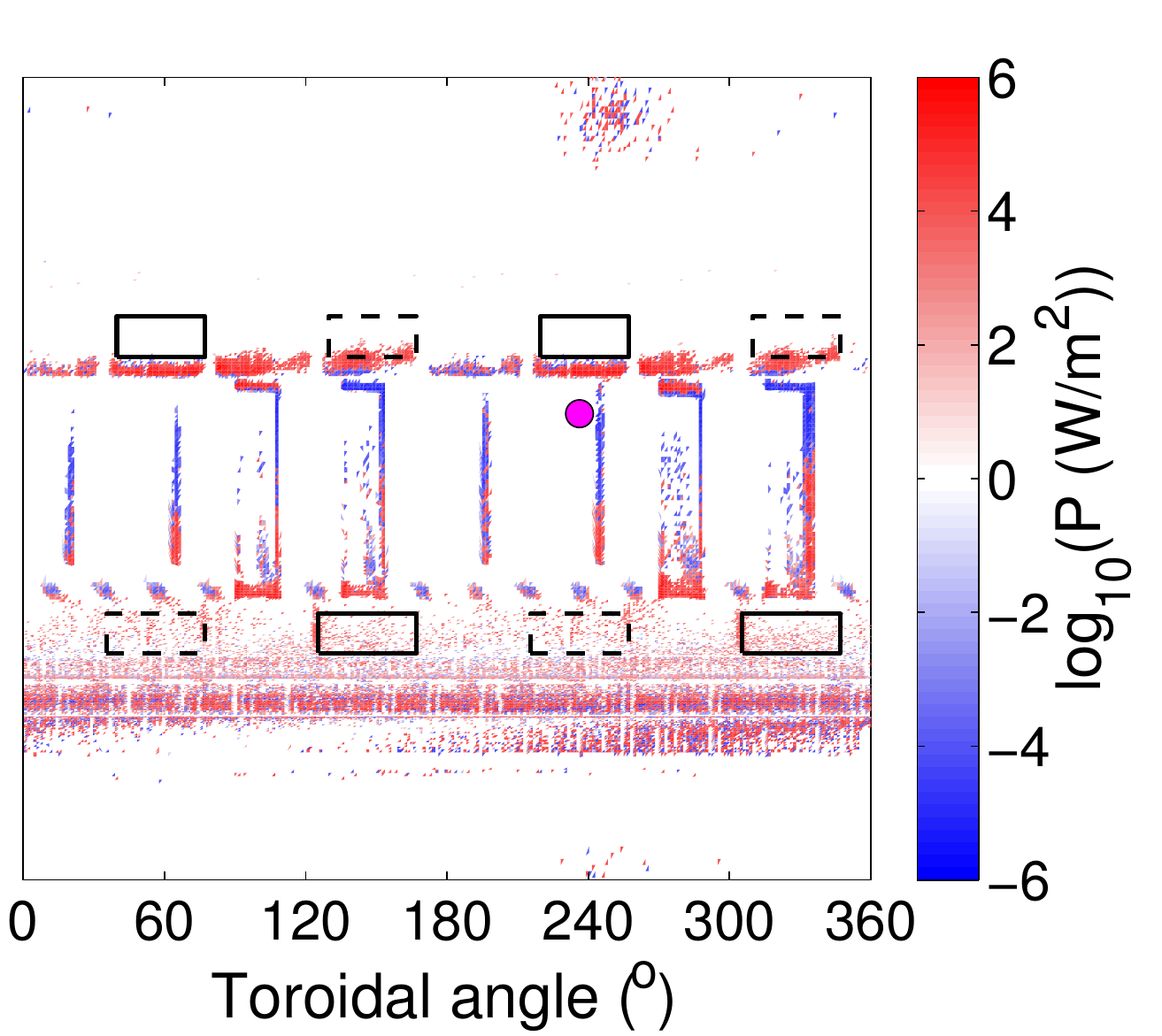}
\hspace{-4.2cm} \raisebox{3.1cm}{\text{(i)}}
\end{minipage}
\caption{Simulated fast ion wall loads. The figures on the left panel ((a), (d), and (g)) illustrate the wall loads for the beams Q5, Q6, and Q8, respectively, when $I_{\textrm{coil}}=0.0$~A, and the figures in the centre panel ((b), (e), and (h)), illustrate the wall loads for the same beams, when $I_{\textrm{coil}}=0.95$~kA. The figures on the right panel ((c), (f), and (i)), on the other hand, illustrate the logarithm of the difference between the two the left and the centre panels. There the increase (decrease) of the wall load in a given region is shown in red (blue). The FILD, located at around 235$^\circ$ in toroidal and 30$^\circ$ in poloidal angle, is marked with a magenta circle. %
The in-vessel coils are drawn as squares with solid (negative current) and dashed (positive current) black line.}
\label{fig:wallLoads}
\end{figure}

\noindent The simulated NBI wall loads for the three neutral beams (Q5, Q6, and Q8) in discharge \#26476 are plotted in Fig.~\ref{fig:wallLoads}; on the left-hand-side column in the absence, and on the right-hand-side column in the presence of the magnetic perturbation created by the in-vessel coils. The effect of the perturbation seems to vary strongly for different beams. For the more perpendicular beams the perturbation increases the losses only slightly (Q5 and Q8 in Figs.~\ref{fig:wallLoads}(a)-(c), and (g)-(i)). On the other hand, for the parallel current drive beam (Q6, Figs.~\ref{fig:wallLoads}(d)-(f)), that has the least losses to begin with, the losses increase drastically. That is, the perturbation has the strongest effect on passing particles. The total power losses without (with) the perturbation are approximately 7\% (9\%), 2\% (9\%), and 4\% (7\%) of the total beam power for the beams Q5, Q6, and Q8, respectively.

For the perpendicular beams Q5 and Q8, protruding wall structures, such as the limiters, collect the majority of the heat load both with and without the magnetic perturbation. For the current drive beam Q6, in addition to the divertor loads that are prominent in all the cases, the majority of the loads are located close to the upper set of coils. This is the case particularly in the presence of the magnetic perturbation. The perturbation also tends to increase the loads right below the coils with negative current and within the coils with positive current.

When using complex 3D magnetic fields, there is always a risk that following only the particles' guiding-centers washes out some of the effects that the magnetic field causes to the real particle orbits. To ensure the validity of the GC simulation results presented above, beam Q6 in \#26476 was simulated (both with $I_{\textrm{coil}}=0.0$~A and $I_{\textrm{coil}}=0.95$~kA) also using the full-orbit following. Because the full-orbit simulations are computationally very expensive, they were done with only 20000 test particles (compared to 800000 test particles used in guiding-center simulations). However, even if the statistics are not very good with so few test particles, total power losses were found to be nearly equal to the GC+FO simulations. Also the locations, as well as the levels, of the peak loads are the same in the two sets of simulations. Hence, the conclusion is that in the simulations performed for this work, guiding-center approximation holds and GC+FO simulations can safely be used without noticeable loss of accuracy.

\begin{figure}[tb]
\begin{minipage}{0.99\linewidth}
\includegraphics[width=0.341\linewidth]{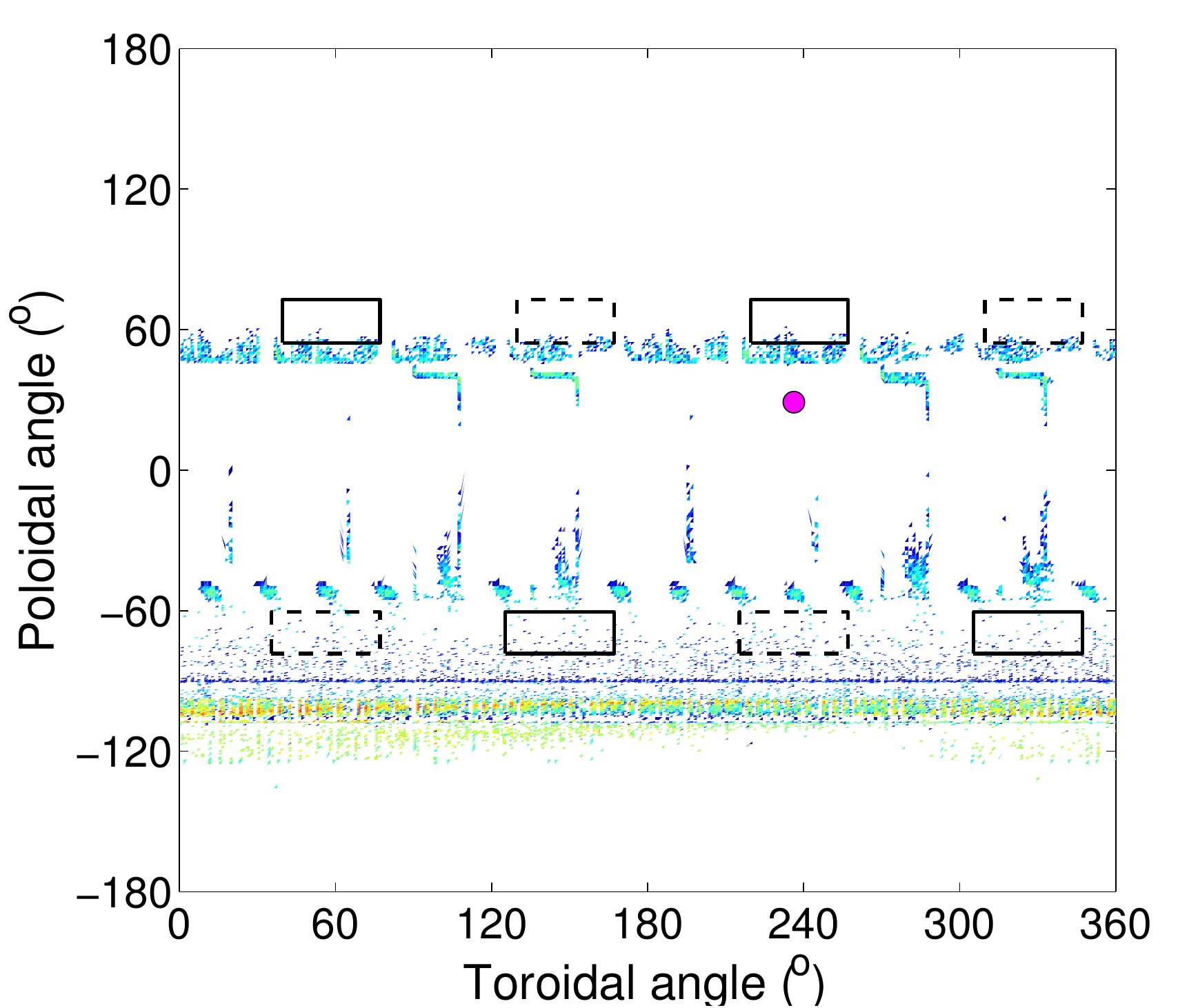}
\hspace{-3.9cm} \raisebox{3.1cm}{\text{(a)}} \hspace{2.9cm} 
\includegraphics[width=0.34\linewidth]{figs/wallload_300_wFILD_noy.pdf}
\hspace{-4.4cm} \raisebox{3.1cm}{\text{(b)}} \hspace{3.4cm} 
\includegraphics[width=0.328\linewidth]{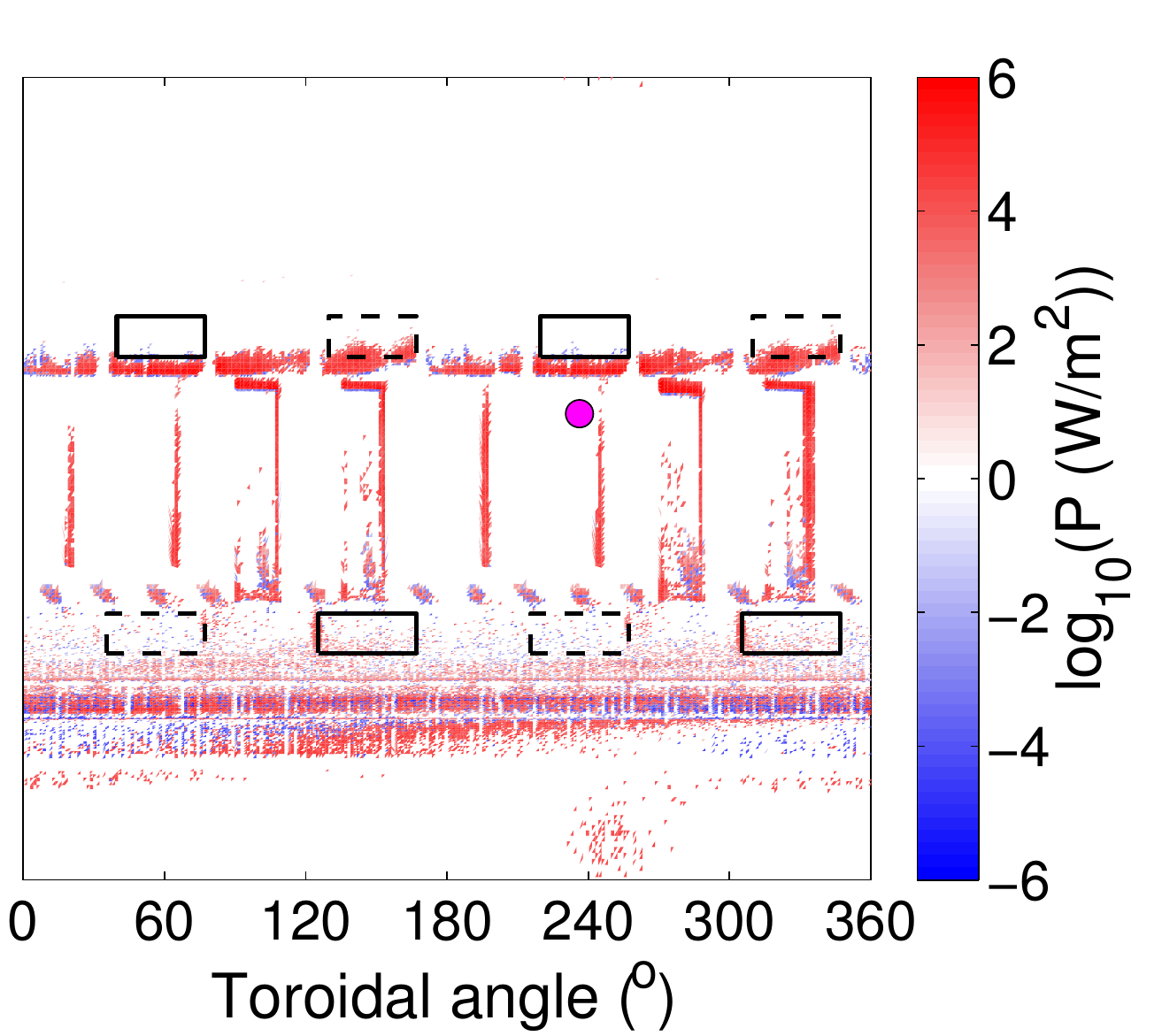}
\hspace{-4.2cm} \raisebox{3.1cm}{\text{(c)}} \\
\end{minipage}
\caption{Simulated fast ion wall loads for the perpendicular beam Q5 in  discharge \#26476 with $I_{\textrm{coil}}=0.95$~kA using (a) pure guiding-center following, and (b) guiding-center following with full-orbit wall collisions. The increase (decrease) of the wall power load on a given element is illustrated in red (blue) in (c). In addition to underestimating the total power losses, pure GC following clearly misses some prominent features in the wall load pattern. For example, the losses on limiters and the hot spots below the coils carrying a negative current (solid black squares) are almost completely disregarded.}
\label{fig:gcWallColls}
\end{figure}

What turns out to be very important, however, is to follow the particles full orbit, instead of its GC orbit, in the vicinity of material surfaces (see Ch.~\ref{sec:ascotSimulations}). In Fig.~\ref{fig:gcWallColls}, the power loads on the walls due to beam Q5 in discharge \#26476 are displayed for both pure guiding-center following (a) and GC following with full-orbit wall collisions (b). The pure GC following produces a qualitatively different deposition pattern, predicting non-existent power loads on the limiters, whereas the GC+FO predicts the limiters to carry significant loads. Pure GC following also underestimates the total power loads, predicting only 2\% of the total NBI power to be lost, compared to 9\% predicted by the GC+FO simulation.

\begin{figure}[tb]
\begin{minipage}{0.99\linewidth}
\includegraphics[width=0.341\linewidth]{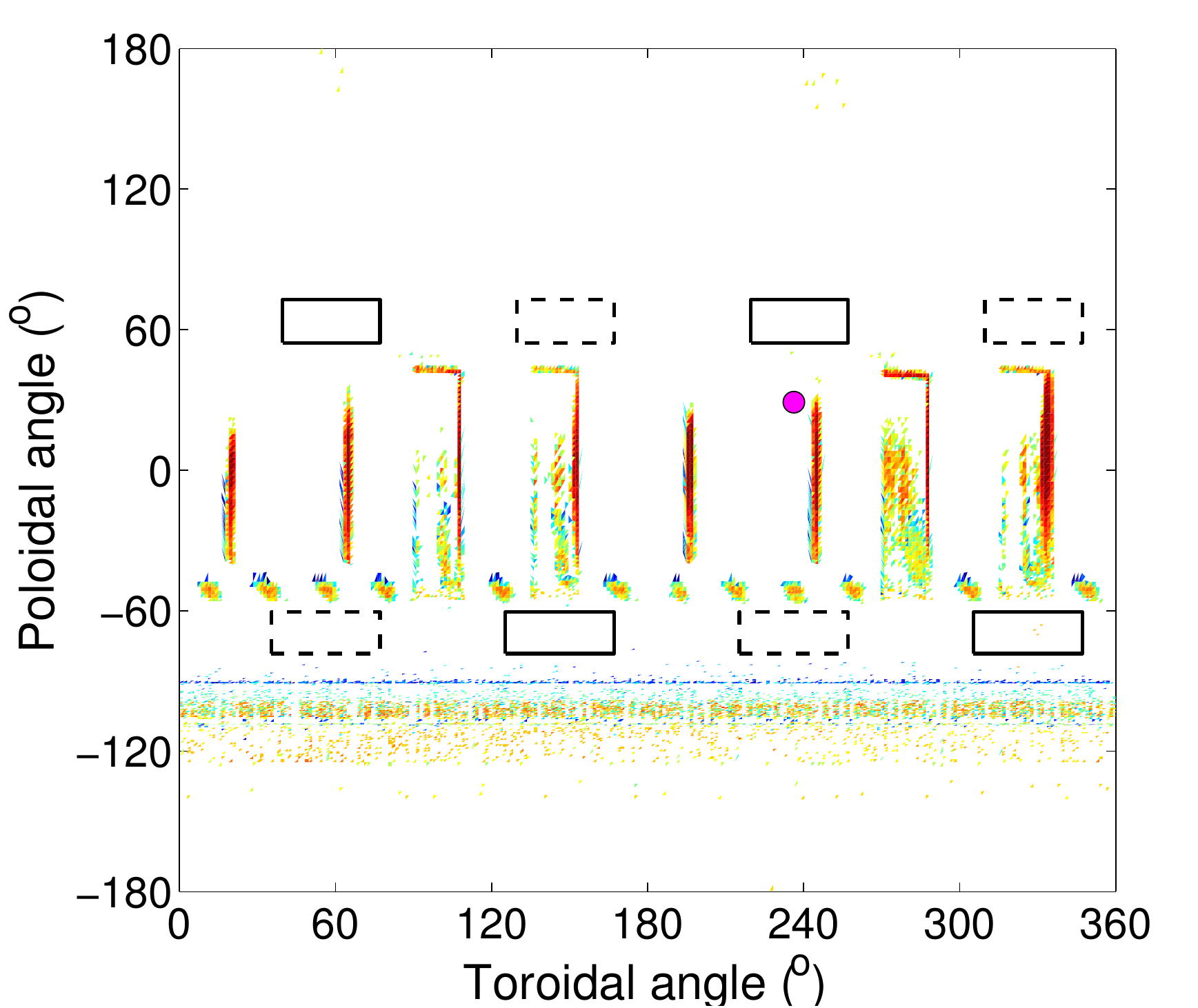}
\hspace{-3.9cm} \raisebox{3.1cm}{\text{(a)}} \hspace{2.9cm} 
\includegraphics[width=0.34\linewidth]{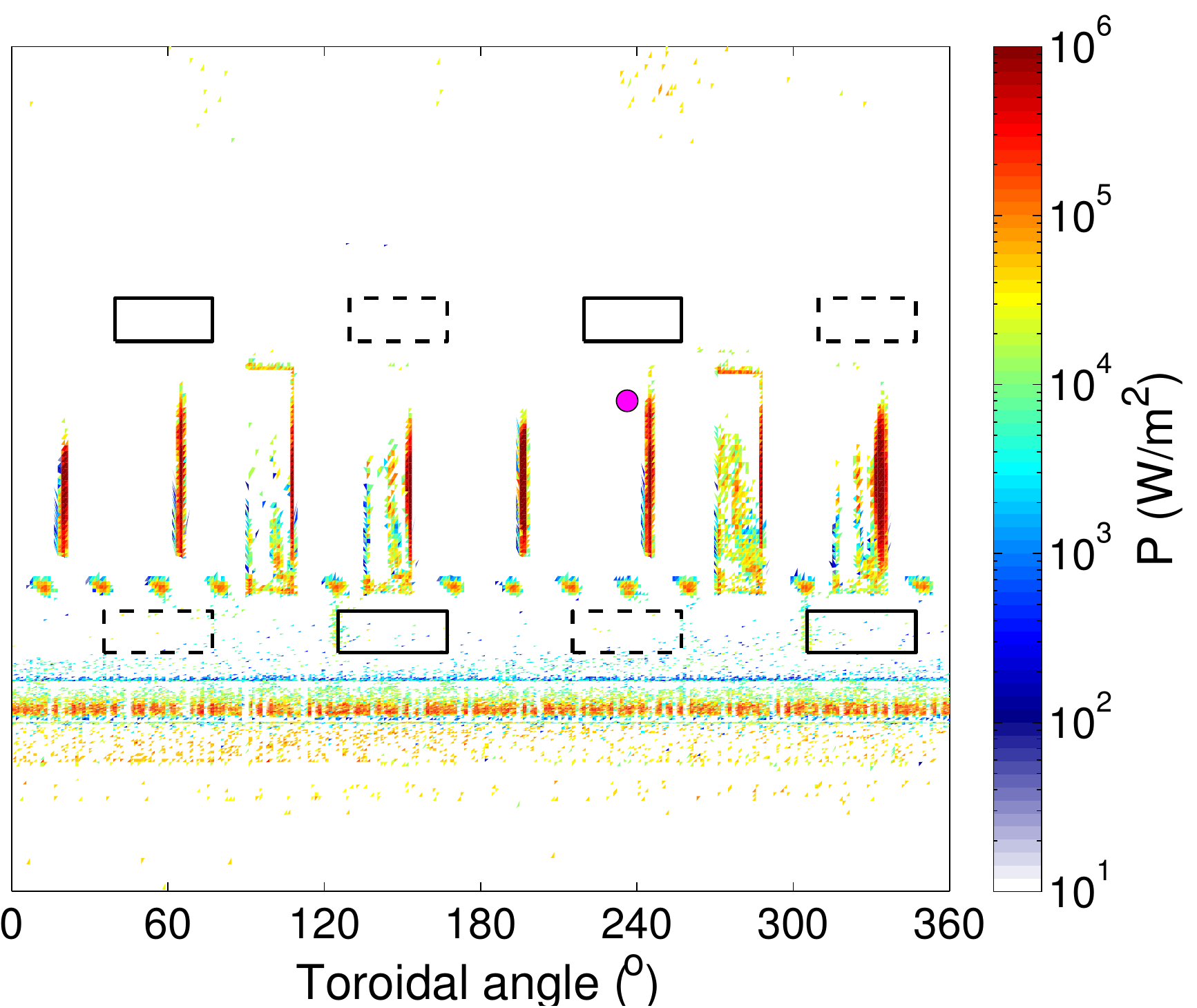}
\hspace{-4.4cm} \raisebox{3.1cm}{\text{(b)}} \hspace{3.4cm} 
\includegraphics[width=0.328\linewidth]{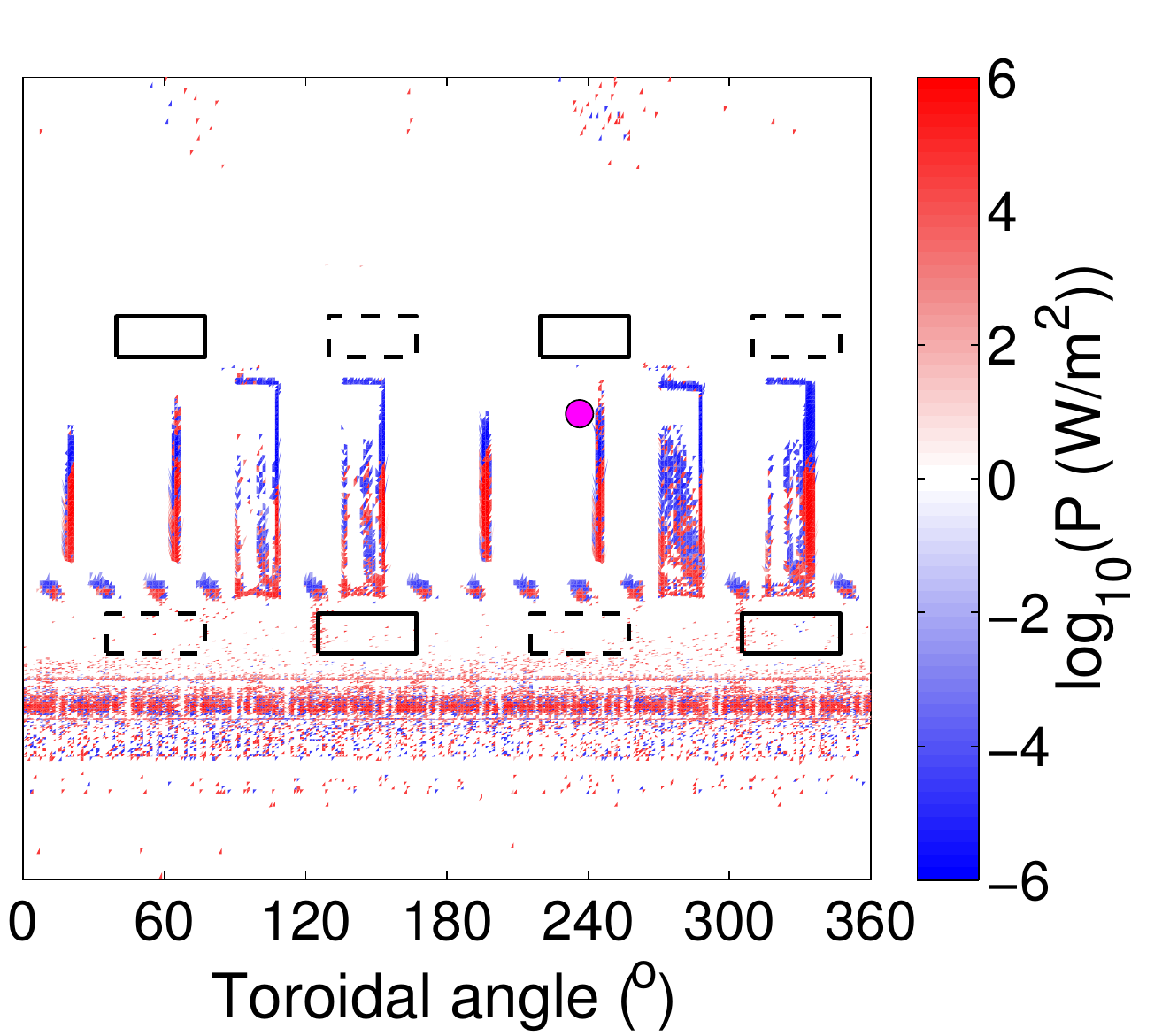}
\hspace{-4.2cm} \raisebox{3.1cm}{\text{(c)}} \\
\end{minipage}
\caption{Simulated fast ion wall loads for AUG discharge \#26895 with (a) $I_{\textrm{coil}}=0.0$~A and (b)  $I_{\textrm{coil}}=0.96$~kA. The increase (decrease) of the wall power load on a given element is illustrated in red (blue) in (c). Running a current in the in-vessel coils merely redistributes the losses increasing the power load on the divertor and reducing it from the limiters while keeping the total lost power at the same level.}
\label{fig:interestingWallLoads}
\end{figure}

In order to study fast ion losses in typical ELM mitigation conditions, neutral beam injected particles were simulated also in discharge \#26895. The combination of beams used in \#26895 was very similar to the one discharge \#26476: one parallel (Q6) and two perpendicular beams (Q3 and Q8). The results, however, are in stark contrast to the ones presented above. In \#26895, the fast particle losses are within statisctical error the same with and without coils. In both cases, 3\% of the total NBI power was lost to the walls. Relatively small differences in the plasma temperatures and densities (cf. Figs.~\ref{fig:26476_tn} and~\ref{fig:26895_tn}) can not explain such a qualitative difference between the two discharges. Instead, it is due to the stronger magnetic field ($B_{\textrm{t}}=2.5$~T in \#26895 compared to 1.8~T in \#26476) smothering the effect of the coils (recall the difference in Figs.~\ref{fig:rippleMap26476} and~\ref{fig:rippleMap26895}). Figure~\ref{fig:interestingWallLoads} illustrates the wall load patterns with $I_{\textrm{coil}}=0.0$~A and  $I_{\textrm{coil}}=\pm 0.96$~kA current running in the in-vessel coils. It shows that, while keeping the total power losses constant, the coils redistribute them. This hypothesis was further confirmed by simulations using the plasma profiles and neutral beams from \#26476 and the strong magnetic field of \#26895. Also in those simulations turning on the coils kept the power losses roughly constant and merely redistributed them. In all cases with a strong magnetic field, turning on the coils increased the power load carried by the divertor while reducing the load on the limiters.

\section{Experimental vs. simulated FILD}
\label{sec:fild}
\noindent FILD measurements are normally dominated by ELMs. Because modelling ELMs is outside the scope of ASCOT and this work, it is important to find the inter-ELM periods and use them as the basis for modelling. Hence, the plasma profiles used in simulations are achieved by averaging the $T_{e}$ and $n_{e}$ measurements over inter-ELM periods. Similar averaging of FILD signal makes it comparable to the simulation results. 

\begin{figure}[tb]
\begin{center}
\includegraphics[width=0.45\textwidth]{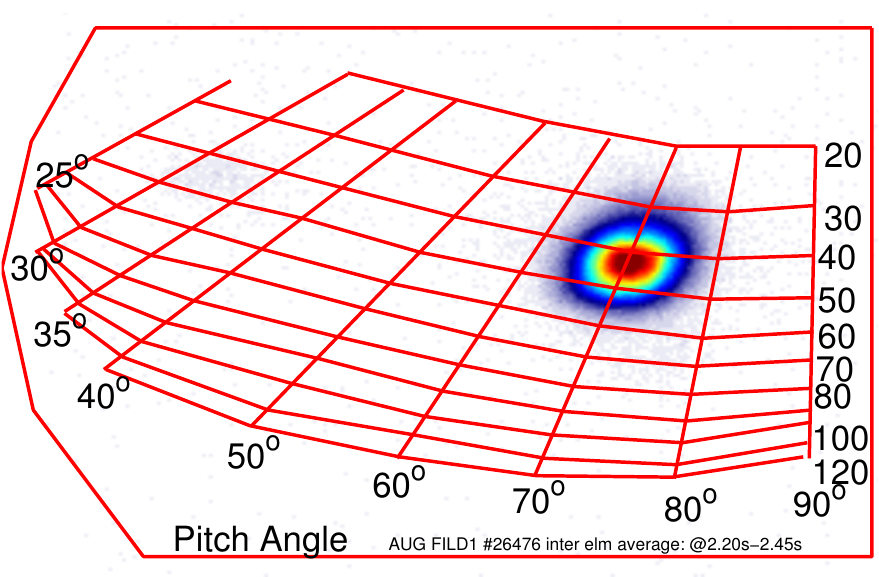}
\includegraphics[width=0.45\textwidth]{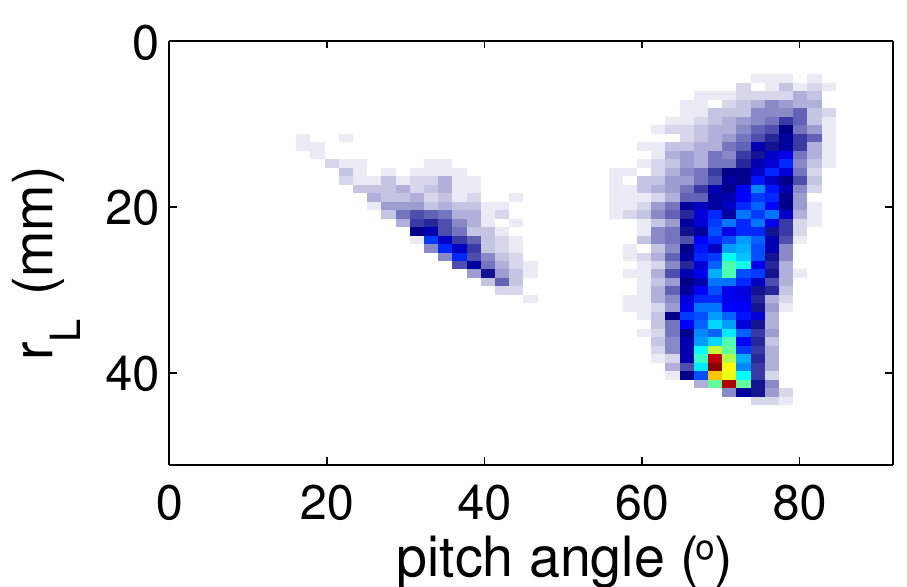}
\includegraphics[width=0.45\textwidth]{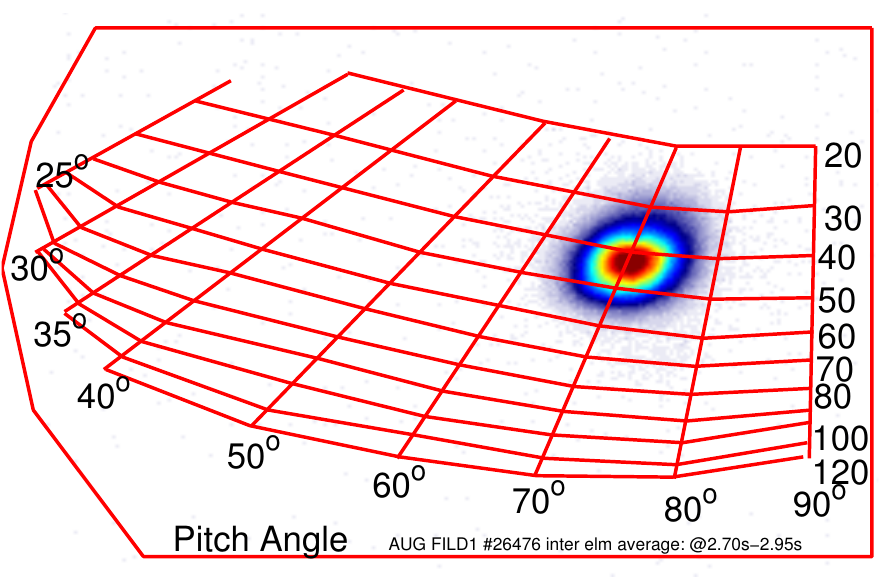}
\includegraphics[width=0.45\textwidth]{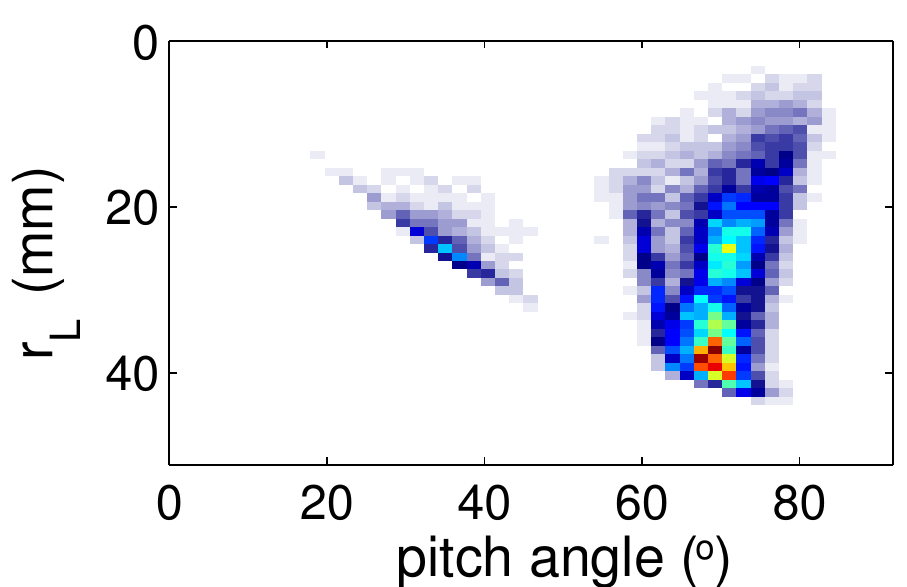}
\caption{Comparison between experimental (left) and synthetic (right) FILD measurements for beam Q5 in discharge \#26476 without (upper) and with magnetic perturbation (lower). Fast ion flux (indicated by color) is in arbitrary units in all the figures. Here pitch angle $\xi = 180^\circ - \arccos{(v_{\parallel}/v)}$.}
\label{fig:fild}
\end{center}
\end{figure}

In the AUG discharge of interest (\#26476), when using Q6 or Q7, ELMs were not completely suppressed but, rather, their amplitude was decreased and frequency increased. This made inter-ELM averaging impossible. When using beam Q5 ELM suppression was not achieved at all and, therefore, good data for experimental versus synthetic FILD diagnostic comparison exists.

The experimental results and the results from a synthetic FILD diagnostic in ASCOT are shown in Fig.~\ref{fig:fild}. There is a good correspondence in both the particle pitch angle ($\xi = 180^\circ - \arccos{(v_{\parallel}/v)}$) and the gyroradius between the measured and the simulated signal; both register highest number of counts at around $\xi=70^\circ$ and $r_{L}=40$~mm. The gyroradii of the particles seen by FILD suggest that most of them are prompt losses. This might, however, not be the case since losses induced by the magnetic perturbation may have similar pitches.
ASCOT synthetic diagnostic, on the other hand, registers a broader distribution of gyroradii (i.e. energy) at around $\xi=70^\circ$. There are several possible reasons for this. The most obvious one is that the experimental FILD rejects detect particles with gyroradii below 20~mm. Another reason might be that particles with small gyroradii do not in reality escape the plasma, but they are simply an artifact in the simulations caused by the vacuum field approximation that overestimates the magnitude and, therefore, the effect of the magnetic perturbations inside the plasma. Since the same approximation is used for calculating both the toroidal field ripple and the magnetic perturbation due to the in-vessel coils, both of the figures in the right-hand panel of Fig.~\ref{fig:fild} would suffer from the same artifact.

Another difference between the experimental and synthetic FILD signals is the second stripe of deposition seen by ASCOT at around $\xi=30^\circ$. This feature is caused by neutral beam particles ionized at the inboard side of the device and promptly (within half a millisecond) lost. Most of it is not visible in the experimental data because losses with pitch angles roughly between 0$^\circ$ and 30$^\circ$ are physically blocked by the FILD collimator or by other protruding first wall structures. Part of the difference could also be due to the synthetic FILD (a cylinder with radius of 0.04~m at $R=2.14...2.24$~m, $z=0.33$~m) being deeper in the plasma than the real one. Judging from these preliminary results it seems that the in-vessel coils have a small effect on the experimental and synthetic FILD signals.
However, there have also been discharges where the experimental FILD signal changes significantly when the coils are turned on. Therefore, more data is needed to isolate the effect of the coils on the FILD signal. Further dedicated experiments are also planned to allow better comparisons between experimental and synthetic FILD.

\section{Conclusions}
\label{sec:conclusions}

The wall power loads caused by neutral beam injected particles were simulated in two ASDEX Upgrade discharges (\#26476 and \#26895) in the presence and absence of the magnetic perturbation induced by the newly installed in-vessel coils. The most recent 3D wall structure of AUG (see Fig.~\ref{fig:aug3d_w_coils}), updated to include the modifications for the 2010--2011 experimental campaign, and 3D magnetic fields were used in the simulations. The toroidal field ripple as well as the magnetic perturbation due to the in-vessel coils were calculated using the vacuum field approximation. That is, plasma shielding, reducing the effect of the perturbation in real experiments, was not taken into account. Consequently, the obtained fast ion wall loads represent the 'worst case scenario' and are likely to be smaller in reality.

For the typical ELM mitigation discharge \#26895, running a current $I_{\textrm{coil}}=\pm 0.96$~kA in the in-vessel coils did not increase the fast particle losses at all. This is due to the stronger magnetic field ($B_{\textrm{t}}=2.5$~T in \#26895 compared to 1.8~T in \#26476) smothering the effect of the coils. For \#26476, where no ELM mitigation was observed, the results were surprisingly different. Three beams were simulated one at a time and turning on the coil current increased the fast ion losses from all three beams. The increase was by far most prominent for the parallel current drive beam Q6 for which the losses were roughly quadrupled (from 2\% to 9\% of total injected beam power). For the perpendicular beams Q5 and Q8, on the other hand, the losses also increased, but not as drastically (from 7\% to 9\% and from 4\% to 7\%, respectively) and the limiters collected a significant fraction of the losses both with and without the magnetic perturbation. In the case of the parallel beam, the losses were more focused around the upper set of in-vessel coils.

The results from Fast Ion Lost Detector (FILD) were in good correspondence with the results of ASCOT synthetic diagnostic. However, in order to better isolate the effect of in-vessel coils on FILD signal, more experiments are needed. A series of dedicated pulses will also be ran to comprehensively benchmark ASCOT against FILD measurements in a well-behaved L-mode plasma.

In the future, fast particles will be simulated in a similar discharge (\#26475), including the observed $\beta$-driven neoclassical tearing mode (NTM) islands. The combined effect of NTM islands and the magnetic perturbation due to the in-vessel coils are expected to increase the number of lost particles and, therefore, the particles seen by the FILD, thus improving the statistics. Another obvious follow-up for this work is to simulate the effect of the ITER ELM-mitigation coils on the confinement and losses of neutral beam injected particles as well as fusion-born alpha-particles.

\vspace{1.0cm}
{\begin{singlespace} \small \noindent {\bf Acknowledgements: } This work was partially funded by the Academy of Finland projects No. 121371 and 134924. This work, supported by the European Communities under the contract of Association between Euratom/Tekes, was carried out within the framework of the European Fusion Development Agreement. The views and opinions expressed herein do not necessarily reflect those of the European Commission. The supercomputing resources of CSC - IT center for science were utilized in the studies.\end{singlespace}}
\normalsize
\section*{References}
\bibliographystyle{otonrevtex}
\bibliography{bibfile}

\end{document}